\documentclass[%
superscriptaddress,
reprint,
 amsmath,amssymb,aps,prx]{revtex4-2}

\usepackage{graphicx}
\usepackage{dcolumn}
\usepackage{bm}
\usepackage{color}
\usepackage[dvipsnames]{xcolor}  
\usepackage{multirow}
\usepackage{ulem}
\usepackage{upgreek}
\usepackage{mathrsfs}

\usepackage[export]{adjustbox}

\usepackage[symbol]{footmisc}

\newcommand{\hl}{}

\graphicspath{{./Figs/}}

\begin{document}


\title{Non-Reciprocal Interactions Induced by Water in Confinement}
\author{Felipe Jim\'enez-\'Angeles}
\affiliation{Department of Materials Science and Engineering, Northwestern University, Evanston, Illinois 60208, United States}
\author{Katherine J. Harmon}%
\affiliation{Applied Physics Graduate Program, Northwestern University, Evanston, Illinois 60208, United States}%
\author{Trung Dac Nguyen}
\affiliation{Department of Chemical and Biological Engineering, Northwestern University, Evanston, Illinois 60208, United States}%

\author{Paul Fenter}
\email{e-mail: fenter@anl.gov}
\affiliation{Chemical Sciences and Engineering Division, Argonne National Laboratory, Lemont, Illinois 60439, United States}%

\author{Monica Olvera de la Cruz}
\email{e-mail: m-olvera@northwestern.edu}
\affiliation{Department of Materials Science and Engineering, Northwestern University, Evanston, Illinois 60208, United States}%
\affiliation{Applied Physics Graduate Program, Northwestern University, Evanston, Illinois 60208, United States}%
\affiliation{Department of Chemical and Biological Engineering, Northwestern University, Evanston, Illinois 60208, United States}%
\affiliation{Department of Physics and Astronomy, Northwestern University, Evanston, Illinois 60208, United States}%
\affiliation{Department of Chemistry, Northwestern University, Evanston, Illinois 60208, United States}%


\begin{abstract}
Water mediates electrostatic interactions via the orientation of its dipoles around ions, molecules, and interfaces. This induced water polarization consequently influences multiple phenomena. In particular, water polarization  affects ion adsorption and transport, biomolecular {self-assembly}, and surface chemical reactions. Therefore,
it is of paramount importance to understand water-mediated interactions modulated by nanoconfinement at the nanoscale.  \hl{Here we investigate the effective interaction between two oppositely charged ions in different positions in water confined between two graphene surfaces. We find that the attraction between physisorbed ions is enhanced in the surface normal direction while the  in-plane interaction is almost unaffected. The attraction in the surface normal direction is further enhanced by decreasing the confinement distance. Conversely, when one ion is intercalated into the graphene layers the interaction becomes repulsive. Moreover, upon exchange
of the ions' positions along the surface normal direction the interaction energy changes by about 5$k_BT$.
The non-equivalent and directional properties found here, referred to as non-reciprocal interactions, cannot be explain by current water permittivity models in confinement. Our X-ray reflectivity experiments of  the water structure near a graphene surface support our molecular dynamics simulation results.}
Our work shows that the water structure is not enough to infer electrostatic interactions near interfaces. \\
Subject Areas: Atomic and Molecular Physics, Chemical Physics, Physical Chemistry
\end{abstract}

\maketitle

\newpage


\section{Introduction}

Water and ions, two basic components of living organisms and ubiquitous in numerous natural and technological processes, are strongly interconnected~\cite{Debenedetti13325,Gotsi_Nat2018,Panganiban1239}. Water mediates interactions between ions, molecules, macromolecules, and interfaces through the polarization of its electric dipole moment. At the same time, charged species induce polarization of other molecules, surfaces, and the medium itself \citep{B917803C}. This leads to complex interactions, the details of which are not yet fully understood. In particular, the capability of a medium to be polarized and the resulting attenuation of electrostatic forces are typically quantified by the dielectric permittivity. This dielectric permittivity normalized by that of the vacuum ($\varepsilon_0$), called the relative permittivity ($\varepsilon_r$), is frequently assumed to be a distinctive property of a material and is often termed as the dielectric constant. Such a denomination, however, is not applicable in nanomaterials, and especially for water in confinement and near interfaces where the properties differ from the bulk \citep{battista2009,acsnano.6b00187,Wu3358,NatCom_2020}.  

\hl{The arrangement of water molecules at interfaces  confers distinctive properties of interfacial water from those in the bulk} \citep{toney94}. \hl{For example, the varying water structure as a function of the distance to an interface modulates ionic interactions} \cite{Chen555}. The ionic interactions near an interface are understood in terms of dielectric permittivity. Hence, considerable efforts have been made to determine the interfacial permittivity~\cite{souza_apl03,Fumagalli1339,Hansen_JCP2005,feller_JCP2003,Netz_la2012,PhysRevE.102.022803} of confined water. These studies have shown that the interfacial permittivity is anisotropic~\cite{Hansen_JCP2005,feller_JCP2003} and is significantly diminished within 1 nm from an interface \cite{Fumagalli1339}. \hl{The connection between this interfacial dielectric permittivity and ionic interactions has been extensively studied using atomistic simulations and continuous models mainly to determine surface ionic profiles} \citep{MO_JCP2015,williams2017,jz200765z} \hl{and ion specific effects} \cite{zhan2019specific,acsanm20}. \hl{Here we study ionic interactions between ions phsysiorbed and intercalated at the water-graphene interace. We show that a dielectric function, even if anisotropic effects due to confinement are included and non-polarizable ions of equal size are considered,  cannot describe the interactions between ions at interfaces (intercalated and physisorbed) and the broken symmetries when the ions are  interchanged (see below).} 

\hl{The graphene/water interface is both technologically relevant and lacks the complexities of heterogeneous surfaces (e.g., proteins), making it an ideal system to investigate fundamental properties of ionic interactions in confinement.} Graphene possesses extraordinary properties such as partial wetting transparency, electrical conductivity, and mechanical strength~\cite{geim2010rise,rafiee2012wetting}. The graphene-water interface is of interest for water desalination \cite{nl3012853,surwade2015water}, for electrochemical energy storage \cite{raccichini2015role} and harvesting \cite{nl2011559}, as a trans-electrode membrane to characterize biomolecules \cite{garaj2010graphene}, and in many other applications. Previous studies at air-water \cite{Perrine13363,HORINEK2009173} and at graphene-water interfaces \cite{McCaffrey13369,jz200765z} have discussed ion adsorption (density profiles) and the effects from ions size, polarizability, \hl {chemical structure (such as OH$^-$ and H$_3$O$^+$)}, and solvation energy, as well as the surface structure, among other factors. \hl{ Here we investigate ion-ion effective interactions near the water-graphene interface. We consider intercalated and physisorbed non-polarizable model ions of equal size but opposite charge. We show that near the interface the interactions between oppositely charged physisorbed ions are non-equivalent by interchanging the ions' positions with respect to the surface and can be shifted from  repulsive to attractive by changing the ionic conformation; that is, when the position of one ion is changed from intercalated to physisorbed. }

%


First, we investigate the water polarization near a graphene surface in the absence of any free charges and observe the changes of the interfacial water polarization in the presence of a nearby ion. Second, we compare the molecular interfacial water structure predicted by the simulation with that determined by X-ray reflectivity measurements. Third, we evaluate the effective ion-ion interactions near the graphene surface both in the plane and along the surface normal and assess the effect of increasing the confinement. We explain our results in terms of the interfacial water polarization rather than the dielectric constant. {We demonstrate that the symmetry breaking of ion-ion interactions is present in both symmetric non-polarizable ion models and in polarizable models of water, ions, and graphene.}  Finally, we analyze our results in the framework of the prevailing continuum theories of electrostatics at interfaces and we demonstrate that continuum theories fail to capture a fundamental breaking of symmetries of ion-ion effective interactions near interfaces. 


\section{Methods} 
\label{sec.methods}

\subsection{Molecular Simulation Models}
\label{ssec.models}
\begin{figure}[!ht]
\includegraphics[trim={0in 0in 0in 0in},clip,width=8cm]{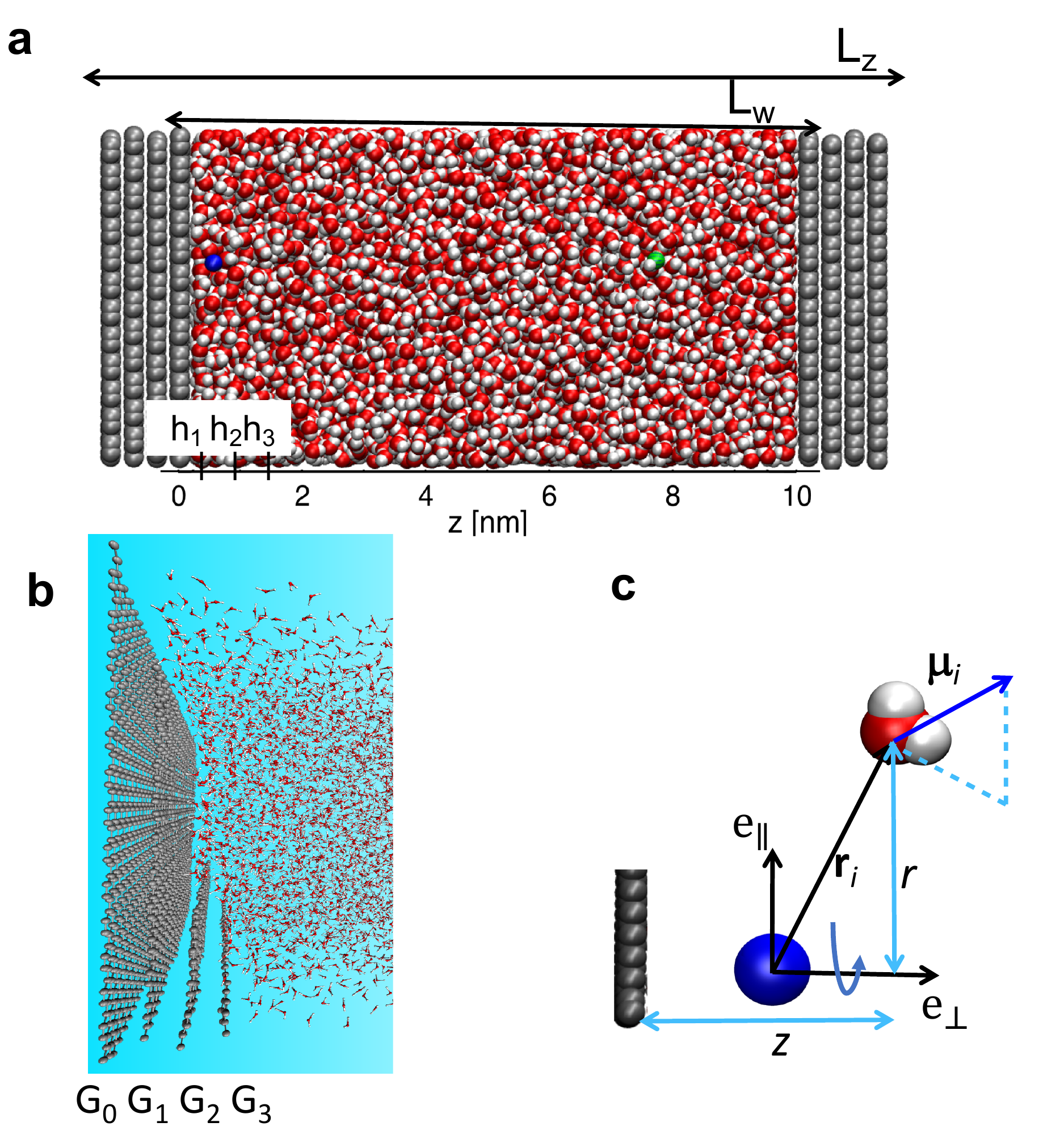}
\caption{{(a)} Simulation setup consisting of a water layer of thickness $L_{\rm w}$ confined between two graphene surfaces. The response of the water polarization to electrical fields is investigated by placing two oppositely charged ions in different configurations. Here the cation (blue sphere) is at about 0.5 nm from the left graphene surface while the anion (green sphere) is at $z\approx 8$ nm; anions and cations have the same Lennard-Jones parameters ($\sigma=0.333$ nm and $\epsilon=1.16\times 10^{-2}$ kJ/mol) and differ only in their valence. $L_{\rm z}$ is the simulation box length in the $z$ direction, and $h_i$ with $i=1,2,3$ are three regions of 0.5 nm employed to quantify the water polarization next to the graphene surface. {(b)} Irregular surface employed for a direct comparison between X-ray reflectivity experiments and MD simulations; the surface is made of four partial graphene layers (G$_0$ - G$_3$) with the relative surface area of each graphene layer determined by the experimental XR results. {(c)} Calculation scheme employed to calculate the polarization field around an ion nearby the water-graphene interface.}
\label{fig0}
\end{figure}

To understand the water-mediated and confinement-altered interactions, we investigate ion-graphene and ion-ion interactions for a pair of non-polarizable \hl{model test ions} {(monovalent anion and cation of equal size)} placed between two uncharged non-polarizable graphene surfaces (see Figure \ref{fig0}a). {These assumptions allow us to show that the phenomena uncovered here are not attributed to specific properties of ions, such  as size and polarizability. Our findings are further confirmed by considering polarizable models of ions, water, and graphene (see Appendix } \ref{sec.polarizability}). The ions are investigated in different configurations with respect to the graphene surface. The simulation box consists of two graphene surfaces separated by a water layer of thickness $L_{\rm w}$; each surface is formed by four graphene layers, $i=0,\dots, 3$, with an inter-layer separation of ~0.358 nm and 1008 carbon atoms per layer. The simulation box size is 5.065 nm $\times$ 5.104 nm $\times$ $L_z$, where the box length along the $z$-direction, $L_z$, is adjusted according to the number of water molecules, $N_{\rm w}$; $N_{\rm w}=$ 8060 and 2233 for $L_{\rm w}=$ 9.8 and 3 nm ($L_{\rm z}=$ 12.28 and 5.63 nm), respectively. The water is simulated using the extended particle charge (SPC/E) model~\cite{spc-e}, and the graphene parameters are taken from the all-atom optimized potentials for liquid simulations (OPLS-AA) force field \cite{ja9621760}. The ions' Lennard-Jones parameters are $\sigma=0.333$ nm and $\epsilon=1.16\times 10^{-2}$ kJ/mol for both ions. Unless explicitly stated, our molecular dynamics (MD) simulations are performed using three-dimensional periodic boundary conditions with slab correction (3DC) to mimic two-dimensional (2D) periodic boundary conditions in the $x$ and $y$ directions \cite{berkowitz}. The system temperature is maintained  at $T=298$ K using a Nos\'e-Hoover thermostat.

To enable a one-to-one comparison between X-ray reflectivity experiments and MD simulations, a separate model was built {with no added ions in the system} and with a graphene surface composed of four layers with different partial coverages (Figure \ref{fig0}b). This mimics the experimental surface wherein the graphene sample is made of multiple incomplete layers. The coverage of each layer was set to the value determined by the experimental X-ray reflectivity best fit structure as follows: $G_0$ contained 1176 carbon atoms with $x \times y$ dimensions 5.157 nm $\times$ 5.955 nm (1 monolayer, ML); $G_1$ contained 924 carbon atoms, $x \times y$ = 5.157 nm $\times$ 4.679 nm (0.78 ML); $G_2$ contained 336 carbon atoms, $x \times y$ = 5.157 nm $\times$ 1.701 nm (0.29 ML); and $G_3$ contained 84 carbon atoms, $x \times y$ = 5.157 nm $\times$ 0.425 nm (0.07 ML). These surface coverages are similar to those determined experimentally.

We analyze our results in terms of the polarization expressed as \cite{jackson1999}
\begin{equation}
 {\bf P}({\bf r}) = {\bf P}_1({\bf r}) - \nabla\cdot {\bf P}_2({\bf r}) + \nabla\nabla\colon {\bf P}_3({\bf r}) +\ldots
 \label{eq:pol}
\end{equation}
which includes the dipole moment per unit volume  ${\bf P}_1$, the quadrupole moment ${\bf P}_2$, octupole moment ${\bf P}_3$, and all higher-order moments. In water, the dipole moment is the main contribution to the polarization. Hence, it can be expressed as 
\begin{equation}
{\bf P}({\bf r})\approx{\bf P}_1({\bf r})=\left\langle \sum_i{\bf \upmu}_i\delta({\bf r}-{\bf r}_i)\right\rangle
\end{equation}
where ${\bf \upmu}_i$ is the dipole moment of the $i$-th water molecule at {\bf r}$_i$, and $\langle \dots \rangle$ represents the ensemble average. The SPC/E water dipole moment is $\upmu_0\equiv${\boldmath$|\upmu_i|$}=0.0489 $e\cdot$nm where $e$ is the positive elementary charge. The calculation scheme in molecular dynamics (MD) simulations is depicted in Figure \ref{fig0}c. Operationally, the instantaneous polarization is calculated as ${\bf p}(r,z)=\frac{\Delta {\bf m}}{\Delta V}$, where $\Delta {\bf m}=\sum_i${\boldmath${\upmu}$}$_i$ includes the dipole moment ({\boldmath${\upmu}$}$_i$) of all the water molecules in the volume $\Delta V$ at $(r,z)$. The mean polarization is calculated as  ${\bf P}(r,z) = \langle {\bf p}(r,z) \rangle$, performing the ensemble average over at least 10$^{4}$ independent configurations. 


\subsection{Potential of Mean Force}
In an $N$-body system the mean force exerted on the $i$-th particle is derived from the instantaneous forces from all the particles and is given by
\begin{equation}
{\bf F}_i(\mathbf{r}^n) = -\nabla_i W^{(n)}(\mathbf{r}^n) = -\langle \nabla_i V(\mathbf{r}^N) \rangle^{(n+1)}
\end{equation}
where $W^{(n)}(\mathbf{r}^n)$ is the $n$-particle ($n\leq N$) potential of the mean force (${\bf F}_i$), $\nabla_i$ is with respect to the coordinates of the $i$-th particle ($i\leq n$); $V(\mathbf{r}^N)=\sum_{i<j}^Nu(\mathbf{r}_{ij})$ is the systems potential energy and $\mathbf{r}^N$ represents the 3$N$ particles' coordinates; $u(\mathbf{r}_{ij})$ is the particles' pair interaction energy; $\langle \dots\rangle^{(n+1)}$ represents the ensemble average over $n+1\dots N$ particles.  The potential of mean force is related to the $n$-particle probability distribution function $\mathscr{G}
^{(n)}(\mathbf{r}^{n})$  by

\begin{equation}
W^{(n)}(\mathbf{r}^{n}) = -k_B T \ln \mathscr{G}
^{(n)}(\mathbf{r}^{n})
\label{pmf}
\end{equation}
In our study $n=2$ and particles 1 and 2 represent the two ions. The potential of mean force  depends on the separation distance between the two graphene surfaces, $L_{\rm w}$, and the ions' positions with respect to the graphene surfaces, $\mathbf{r}_2$ and $\mathbf{r}_1$;  $\mathbf{d}=\mathbf{r}_2-\mathbf{r}_1$ is the ions' relative position and $d=|\mathbf{r}_2-\mathbf{r}_1|$. We investigate the following cases: 
\begin{enumerate}
    \item $W(z)=W(\mathbf{r}_1,\mathbf{r}_2)$ is the interaction of one ion, say 1, and the graphene surface along the surface normal direction, where $|\mathbf{r}_1|=z_1\lesssim 1.5$ nm, $|\mathbf{r}_2|\to L_{\rm w}$; $L_{\rm w} \approx 10$ nm. The interaction of ion 2 with the surface is calculated by exchanging the ions' position.
    \item $W_\bot(d)=W(|\mathbf{r}_1-\mathbf{r}_2|)$ is the ion-ion interaction along the graphene  surface normal direction when $x_1=x_2$, $y_1=y_2$, and $z_1=0.28$ nm and $z_2\lesssim 1.5$ nm, or by exchanging $z_1$ and $z_2$.
    \item  $W_{||}(d)=W(|\mathbf{r}_1-\mathbf{r}_2|)$ is the ion-ion interaction in the plane parallel to the graphene  surface when $z_1=z_2\approx 0.28$ nm,
    \item  $W_{b}(d)=W(|\mathbf{r}_1-\mathbf{r}_2|)$ is the ion-ion interaction in bulk (i.e., with no the graphene surfaces).
    
\end{enumerate}

To calculate the potential of mean force (PMF) we use the umbrella sampling method \cite{TORRIE1977187} which is based on  Eq. \eqref{pmf}.  The technique consists in performing biased sampling by fixing the ion of interest at designated positions along the reaction coordinate $\xi = z,$ $d$. The test ion is fixed using a harmonic potential given as $u_{umbrella}(\xi)=-k(\xi-\bar{\xi}_i)^2$ where $\{\bar{\xi}, i=1,\dots, M\}$ is a set of $M$ equilibrium positions and $k=$ 2000 kJ/(mol nm$^2)$ is the spring constant. The separation distance between two contiguous $\bar{\xi}_i$ and $\bar{\xi}_{i+1}$ equilibrium position is 0.03 nm, approximately. The system is simulated  over at least 20 ns to generate a distribution functions around for each $\bar{\xi_i}$. The biased potential of mean force is related to the particle's distribution functions within the windows by Eq. \eqref{pmf}. The unbiased PMF, the biased PMF, the biased probability distribution function, and the external potential are related in an exact way \cite{wcms.66}. The weighted histogram analysis method (WHAM)  is employed to construct the unbiased PMF \cite{Hess08}. 


\subsection{Continuum Theory of Electrostatics}
\label{ssec:conti}
In the classical theory of electrostatics, an interface is modeled by taking into account two media of dielectric constants $\varepsilon_1$ and $\varepsilon_2$, respectively \cite{jackson1999}. The dielectric mismatch between the two media leads to the following boundary conditions for an electric field {\bf E} passing through the interface
\begin{eqnarray}
 \varepsilon_1 E_\bot^{(1)} &=& \varepsilon_2 E_\bot^{(2)} \nonumber \\
 E_{||}^{(1)} &=& E_{||}^{(2)} 
 \label{boundary_c}
\end{eqnarray}
where $E_\bot$ and $E_{||}$ represent the electric field components perpendicular and parallel to the interface, respectively, passing through medium 1 or 2 as indicated by the superscripts. The boundary conditions in Eq. \eqref{boundary_c} imply that the work needed to bring a particle of charge $q$ from infinity to a distance $z$ from an interface of planar geometry is given by   

\begin{equation}
W_{\rm c}(z)=\frac{\alpha q^2}{4\pi\varepsilon_0\varepsilon_{1} z}
\label{image_w}
\end{equation}
where the dielectric mismatch is {quantified by $\alpha=(\varepsilon_{1}-\varepsilon_{2})/2(\varepsilon_{1}+\varepsilon_{2})$}  \citep{NGUYEN201980,Zwanikken5301}.

\subsection{Molecular theory of permittivity}
\label{ssec:molect}
In the molecular theory of dielectrics, the interfacial region is described by means of a local permittivity {\boldmath ${\varepsilon}$}~\cite{Hansen_JCP2005} which is a function of the system's molecular parameters. The components of the permittivity tensor are calculated from the {\it unperturbed} interfacial water structure, i.e., in the absence of free charges within the interfacial region (See appendix \ref{sec.permittivity}). In a system {with} slab geometry (see Figure \ref{fig0}a), the permittivity tensor is a diagonal matrix with components ${\varepsilon}_{||}(z)\equiv{\varepsilon}_{xx}(z)={\varepsilon}_{yy}(z)$, and ${\varepsilon}_{\bot}(z)\equiv{\varepsilon}_{zz}$. 

\begin{equation}
\varepsilon_{||}(z)=1 +(\frac{\beta}{2V\varepsilon_0})(\langle {\bf M}_{||}\cdot{\bf p}_{||}(z)\rangle-\langle {\bf M}_{||}\rangle\cdot\langle{\bf p}_{||}(z)\rangle)
\label{eqn1}
\end{equation}
and for the inverse permittivity $\varepsilon^{-1}_{\bot}(z)$ in the direction perpendicular to the surface

\begin{equation}
 {\varepsilon^{-1}_{\bot}(z)}=1-(\frac{\beta}{V\varepsilon_0})\frac{(\langle {\bf M}_{\bot}\cdot{\bf p}_{\bot}(z)\rangle-\langle {\bf M}_{\bot}\rangle\cdot\langle{\bf p}_{\bot}(z)\rangle)}{D_{\bot}}
\label{eqn2}
\end{equation}
where  $||$ and $\bot$ indicate, respectively, the parallel and perpendicular components of the total dipole moment {\bf M} and the instantaneous polarization ${\bf p}(z)$, $\langle\ldots\rangle$ means the ensemble average, $\beta=1/(k_BT)$,  $T$ is the absolute temperature, $k_B$ is the Boltzmann constant, $\varepsilon_0$ is the vacuum permittivity, and $V$ is the simulation box volume; 

\begin{equation}
D_{\bot}=1+(\frac{\beta}{V\varepsilon_0})\int \left[\langle {\bf M}_{\bot}\cdot{\bf p}_{\bot}(z)\rangle-\langle {\bf M}_{\bot}\rangle\cdot\langle{\bf p}_{\bot}(z)\rangle\right] dV 
\end{equation}

Equations \eqref{eqn1} and \eqref{eqn2} are applicable in systems where 3D periodic boundary conditions are assumed ($x$, $y$, and $z$ directions). Ballenegger and Hansen \cite{Hansen_JCP2005} derived the expression for 2D periodic boundary ($x$ and $y$ directions) where $D_{\bot}=1$. The results from simulations using 3{D} and 2{D} boundary conditions are quantitatively different.

\subsection{Experimental Methods}

\begin{figure}[!h]
    \centering
    \includegraphics[trim={0in 0in 0in 0in},clip,width=7.5cm]{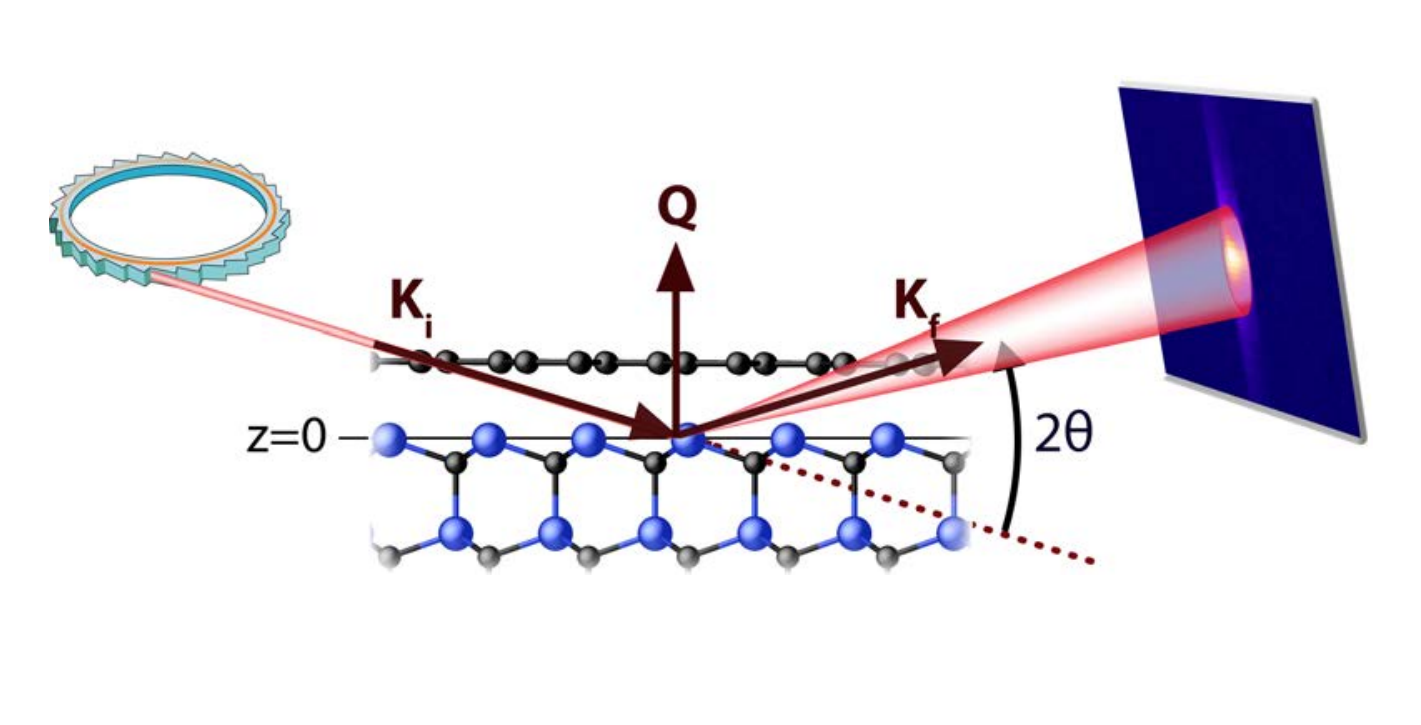} 
    \caption{The specular X-ray reflectivity (XR) is measured as a function of the perpendicular momentum transfer Q = $\lvert{\bf K}_f$-${\bf K}_i\rvert$ where the incident ${\bf K}_i$ and reflected ${\bf K}_f$ X-ray beams vary by the scattering angle 2$\uptheta$.}
    \label{fig:exp_md}
\end{figure}

High-resolution X-ray reflectivity (XR) experiments provide a sensitive probe of the molecular structure of well-defined interfaces via a direct measurement of the electron density distribution~\cite{RENAUD19985} and have been employed to study the interfacial water structure at numerous planar surfaces \cite{fenter_lee_2014,Hass2008,Emery2013,Zhou2012}. Such measurements then serve as an indirect probe of the molecular dipole orientation and polarization at the interface. Specular XR measurements of the graphene/water interface, i.e., along the graphene (0001) surface normal direction, were carried out using a 3 mm $\times$ 10 mm sample of an epitaxial graphene thin film grown on a semi-insulating 6H-SiC(0001) substrate (EG/SiC) immersed in ultra-pure de-ionized water (DIW, nominal pH = 7). EG was grown by thermal decomposition of SiC under a partial pressure of Ar at 1500 $^\circ$C according to methods previously described~\cite{EGgrowth2009,deHeer16900}. This approach produces incomplete graphene layers as a growth artifact. Nevertheless, the resulting sample with large-area graphene surfaces is better suited for XR measurements than mechanically exfoliated graphene, which produces micron sized flakes. Measurements were performed at beamline 33-ID-D of the Advanced Photon Source at Argonne National Laboratory. A photon energy of 14 keV ($\lambda =$ 0.89 {\AA}) was used, and the X-ray beam cross section measured approximately $\sim50$ $\mu$m $\times$ 1 mm ($h \times w$). The specular XR data probe the time- and laterally-averaged (in the $xy$-plane) interface structure as a function of the vertical momentum transfer, $Q = 4\pi \sin(2\theta/2)/\lambda$ where $2\theta$ is the X-ray scattering angle (see Figure \ref{fig:exp_md}). The sample area probed by the X-ray beam varies with the scattering angle and ranges from $\sim1.58$ mm$^2$ to $\sim0.14$ mm$^2$ for the present measurements.

The XR best fit structure was derived following a non-linear least squares optimization procedure. In general, the electron density distribution of each atomic layer, \textit{j}, along the substrate surface normal direction is modeled as a Gaussian with parameters to describe its position, $z_j$, coverage, $\theta_j$, and thermal widths, $u_j$. The total electron density profile is given by 
\begin{equation}
    \label{eq:rho_e}
    \rho(z)=\sum_{j}\frac{Z_j\Theta_j}{\sqrt{2\pi u_j^2}}e^\frac{-(z-z_j)^2}{2u_j^2}
\end{equation}
where $Z_j$ is the atomic number of the $j^{th}$ layer. The reflectivity signal $R(Q)$ is related to the modulus squared of the Fourier transform of the electron density distribution and can be calculated for any set of model input parameters as
\begin{equation}
    R_{calc}(Q)\propto\left|\sum_{j}f_{0,j}\Theta_je^{iQz_j}e^{-Q^2u_j^2/2}\right|^2.
\end{equation}
Here, the proportionality indicates that experimental considerations such as the angle-dependent transmission of X-rays through the sample cell and surface roughness must be accounted for as extrinsic factors in the model (see Appendix C); $f_{0,j}$ is the atomic scattering factor of each element $Z_j$ \cite{IUCr} and accounts for the $Q$-dependent decay of the X-ray scattering intensity resulting from the spatial distribution of electrons around an atomic core. We evaluate the accuracy of the model via the goodness-of-fit metric $\chi^2$ 
\begin{equation}
    \label{eq:chi2}
    \chi^2=\frac{1}{N}\sum_{Q}\left(\frac{R_{calc}(Q)-R_{exp}(Q)}{\sigma(Q)}\right)^2
\end{equation}
where $N$ is the number of data points, $R_{exp}(Q)$ is the experimentally measured reflectivity, and $\sigma(Q)$ is the experimental uncertainty at $Q$. The model parameters are refined until $\chi^2$ converges ($\chi^2=1$ for a perfect fit within experimental uncertainties). 

The model for the present system consists of a semi-infinite bulk 6H-SiC(0001) substrate~\cite{SiCStruct}, one unit cell of SiC (i.e. six alternating C-Si layers) to describe surface, up to eight C layers with the density of 2D graphene, and a semi-infinite layered water model described by Magnussen \textit{et al.}~\cite{Magnussen_1995}. The interfacial SiC, graphene, and water structures are optimized according to Eqs. \eqref{eq:rho_e}-\eqref{eq:chi2} while the bulk SiC structure is fixed. 

Given a single adsorption surface (i.e., for a uniform, complete graphene layer), the layered water model includes a series of $m$ Gaussians ($m=0,1,2...$ with the zeroth layer being closest to the adsorption surface) along the surface normal. The position $z_m$ and width $u_m$ of each layer are given by 
\begin{eqnarray}
    z_m &=& z_0 + md_w \\
    u_m &=& \sqrt{u_0^2+m\bar{u}^2}
    \label{eqn6}
\end{eqnarray}
where $z_0$ is the height of the zeroth Gaussian relative to the substrate surface (i.e., the interfacial water height), $d_w$ is the distance between adjacent Gaussian peaks, $u_0$ is the width of the zeroth layer, and $\bar{u}$ is the width broadening of subsequent layers such that the density asymptotically approaches that of bulk water ($\rho_w= 330$ $e^-$/nm$^3$, where $e^-$ is the negative elementary charge). We assume that the areal density of water in each Gaussian of this layered water distribution has the same coverage due to a lack of confinement in the lateral directions in our system. The coverage of each layer is then given by
\begin{equation}
    \Theta_w=\frac{A_{UC}d_w}{V_w}
    \label{eqn7}
\end{equation}
where $A_{UC}$ is the unit cell area of the SiC substrate and $V_{\rm w}=0.0299$ nm$^3$ is the effective volume of a water molecule in bulk assuming spherical symmetry. 

As noted previously, the EG/SiC growth methodology used in this work leads to partial layers of graphene. Therefore, we modified the layered water model \cite{Zhou2012} to incorporate multiple graphene surfaces, $G_n$ ($n=0,1,2,...$), and assumed that water interacts in the same way with each graphene layer. Namely, above each exposed graphene surface exists the same intrinsic water structure according to Eqs. \eqref{eqn6} and \eqref{eqn7} but with a modulation due to the position $z_n$, width $u_n$, and coverage $\Theta_n$ of the adsorbing graphene layer. All together, the layered water structure is described by
\begin{eqnarray}
 z_{m,n} &=& z_m + z_n\\
 u_{m,n} &=& \sqrt{u_m^2+u_n^2}\\
 \Theta_{w,n} &=& \Theta_w \frac{\Theta_{n}-\Theta_{n+1}}{\Theta_{\textrm{ML}}}
 \label{eqn8}
\end{eqnarray}
where the graphene monolayer coverage on SiC is $\Theta_{\textrm{ML}}=$ 3.147 carbon atoms per SiC unit cell. Because the least-squares fitting finds a local minimum in the parameter space, multiple structures consistent with the same XR data are possible. Therefore, we constrained the model parameters of the SiC and graphene surfaces based on previous chemically-resolved measurements of the air/EG/SiC interface \cite{Emery2013}. Further details of the XR analysis and all best-fit parameter values are reported in Appendix C.


\begin{figure*}
\includegraphics[trim={1.5in 3.8in 1.5in 1.1in},clip,width=12cm]{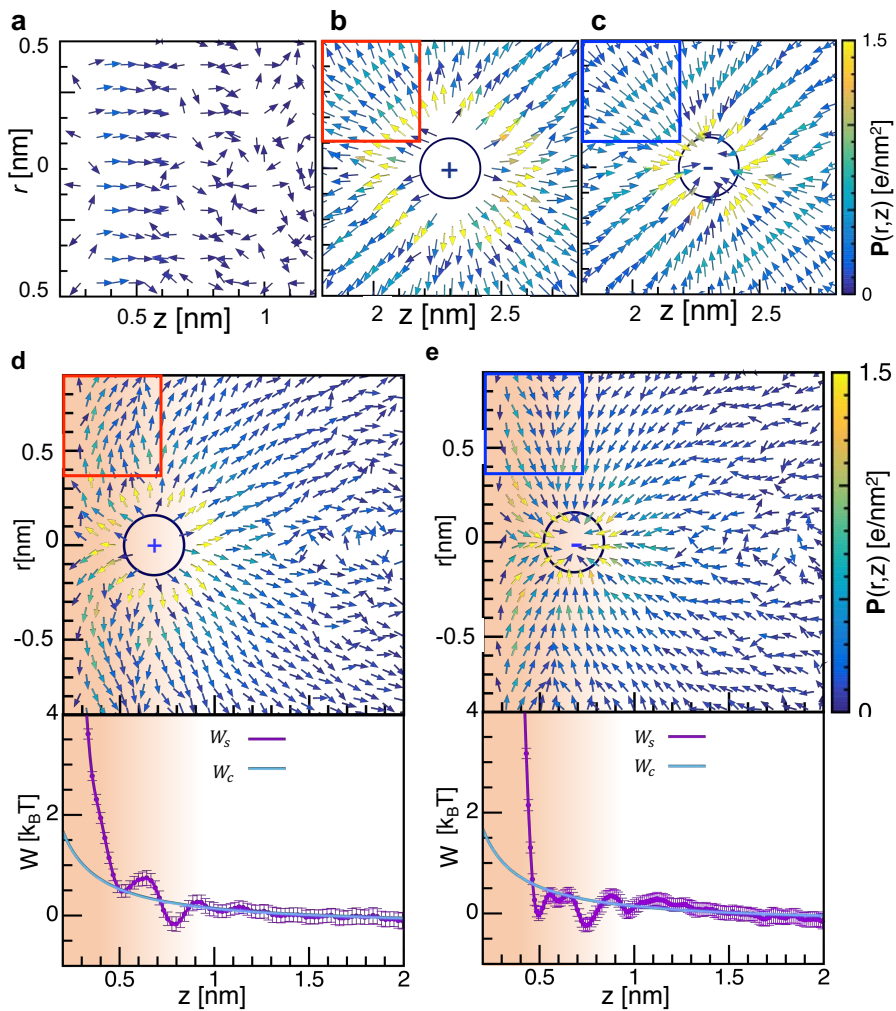}
\caption{Water polarization near the graphene surface and around ions.  When both ions are away from each other and away from the graphene surface ($z_1\approx 2.3$ nm and $z_2\approx 8$ nm), an intrinsic polarization is observed {(a)} near the graphene surface, {(b)} around the cation, and {(c)} around the anion; the arrows indicate the polarization orientation and the magnitude is given by the scale bar at the right;  $r$ is the distance from the ion center along the direction parallel the graphene surface (${\bf e}_{||}$) while $z$ is the distance from the graphene surface in the perpendicular direction (${\bf e}_{\bot}$). The water polarization near the graphene surface is changed by the presence of a nearby {(d)} cation or {(e)} anion. The bottom panels show the ion potential of mean force profile to bring an ion from the bulk to the graphene surface from our molecular dynamics simulations $W_{\rm s}$ (purple line) and calculated via Eq. \eqref{image_w} form the continuum  theory of electrostatics $W_{\rm c}$ (light-blue line). \hl{The boxes in b to e show the regions of the polarization maps that are different in bulk and in the presence of a physisorbed ion;  red (b and d) corresponds to the cation and blue (d and e) corresponds to the anion.}}
\label{fig1}
\end{figure*}

\section{Results}

\subsection{Interfacial water polarization}

The polarization field of water near the graphene surface and in the absence of nearby test charges is shown in Figure \ref{fig1}a. Even if the surface is uncharged and there are no nearby ions and no external electric field, the water is polarized perpendicularly to the graphene surface. This intrinsic interfacial polarization is due to the water dipole moment alignment caused by the solid surface {and is also observed when polarizable models of water and graphene are used in the simulations (see Appendix} \ref{sec.polarizability}). This preferential orientation decays in an oscillatory way to a random polarization in the bulk ($z>1$ nm; see Fig. \ref{fig1}a). As a reference for ions in bulk we show the polarization field around a cation and around an anion far away ($z>$ 2 nm) from the interface in Figures \ref{fig1}b and \ref{fig1}c, respectively. In this case, the test charges are sufficiently far from the graphene surface and do not perturb the interfacial properties. At this separation distance from the graphene surface the water polarization around the ions is spherically symmetric and decays without oscillations. However, the polarization field is not spherically symmetric when one ion is placed within $\sim$1 nm of the graphene surface (see Figures  \ref{fig1}d and \ref{fig1}e for the nearby cation and anion, respectively) while the oppositely charged ion is kept away ($z\approx 8$ nm).  Moreover, the polarization response is asymmetric and unequal around the anion and cation. \hl{For example, the polarization of water around a cation in bulk is outward and radially symmetric whereas the component of the polarization in the $z$-direction is reversed in the presence of graphene (see the red boxes in Figs.} \ref{Fig3}b,d). \hl{The polarization around an anion in the bulk is inward and radially symmetric whereas near a graphene surface the polarization is aligned parallel to the graphene surface (see the blue boxes in Figs.} \ref{Fig3}c,e). This result contrasts with the continuum theory of electrostatic where the polarization fields for positive and negative test ions are assumed to have the same magnitude and opposite direction.

The unequal responses are quantified by the potential of mean force (PMF), $W(z)$, and is different for the anion and the cation (even if both have the same radius). For example, the cation's PMF at $z\approx 0.5$ nm is approximately 0.5 $k_BT$ while the anion's PMF is approximately 0 $k_BT$. Additionally, for $z\lesssim 0.5$ nm the PMF becomes steeper for the anion than for the cation. The interfacial ion specificity has been attributed to the ionic polarizability, size, and valence. However, our results show that solely a change in sign gives rise to a pronounced ion specificity via the asymmetric water polarization response to the sign of the ions. In continuum electrostatic theory the change in the dielectric permittivity at the interface is modeled by the image charge method, which from water to graphene would assign an equally repulsive force for the negative and positive ions~\cite{jackson1999} (see light-blue line in Figs. \ref{fig1}d,e). This symmetry is also obtained in continuum models for electrolytes in confinement \cite{Zwanikken5301,santos2015,PRE.70.051802}.

\subsection{Interfacial Water Electron Density Distribution}

\begin{figure*}
\includegraphics[trim={0in 0in 0in 0in},clip,width=12cm]{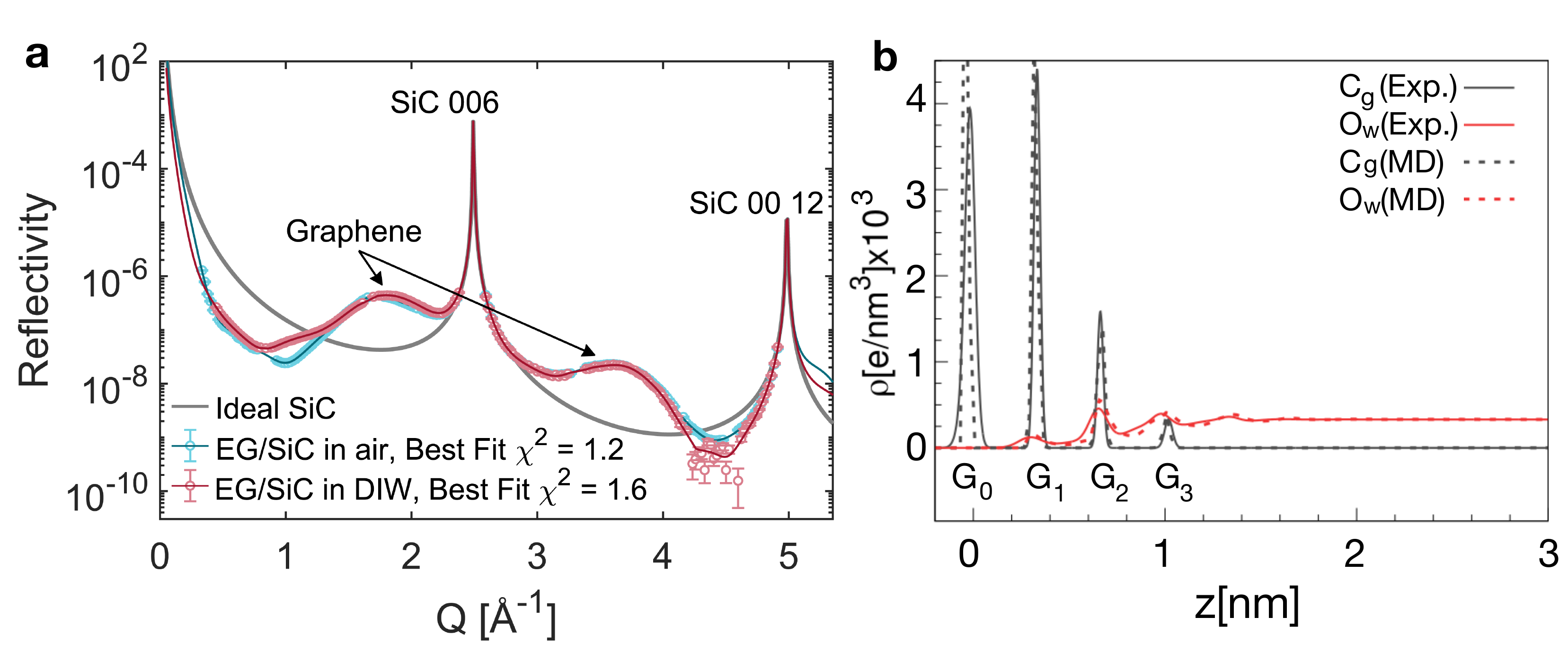}\\
\includegraphics[trim={0.2in 0in 0.2in 0in},  clip,width=12cm]{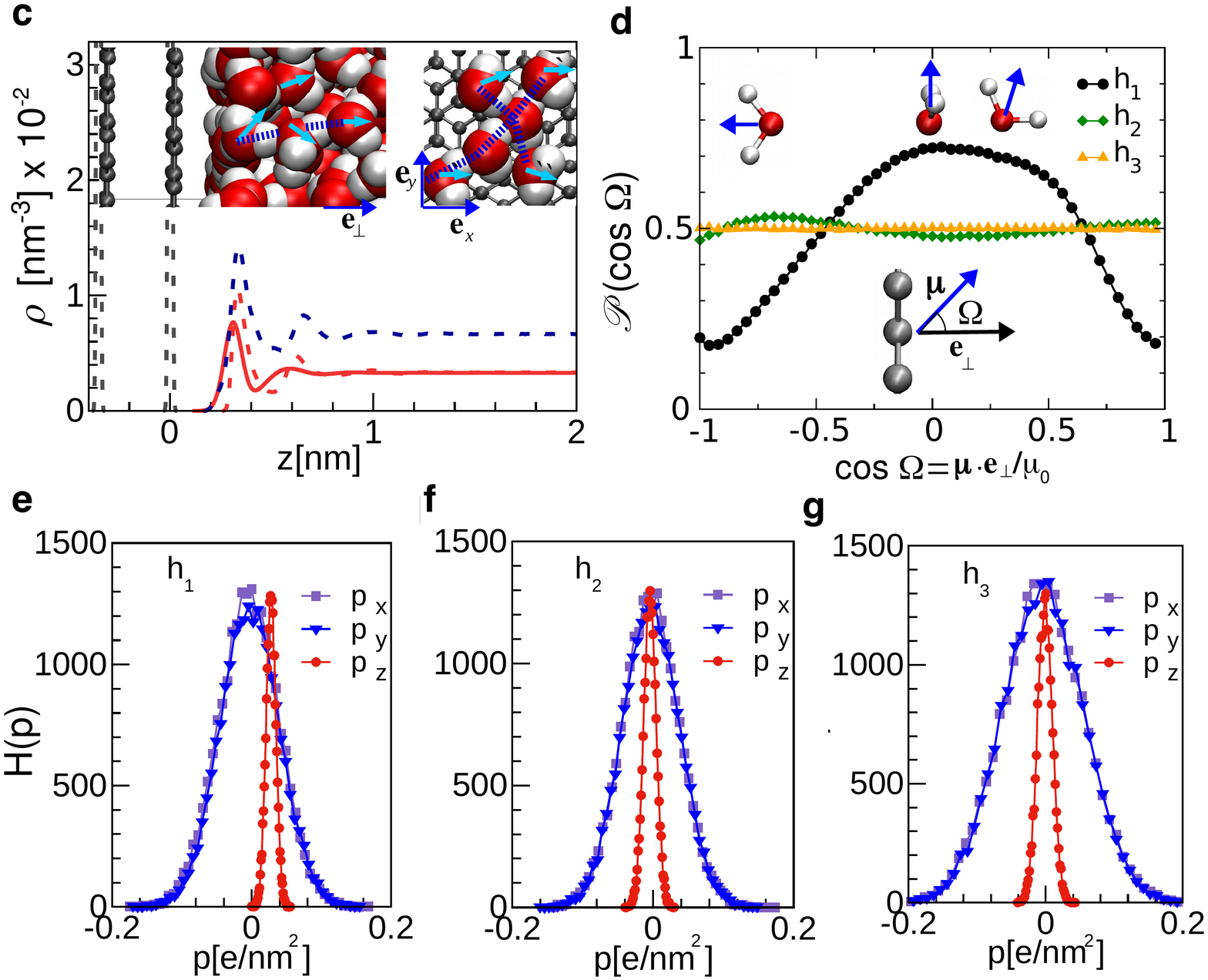}  
\caption{The graphene-water interface from experiments and molecular simulations. (a) \hl {The measured XR intensities of EG/SiC in air (blue circles) and EG/SiC in deionized water (red circles) are shown with their best fits, $\chi^2 =$ 1.2 in air (blue line) and $\chi^2 =$ 1.6 in DIW (red line), and the expected intensity from an ideally-terminated SiC substrate for reference (gray line); the Bragg peaks of the SiC substrate and graphene film are indicated.}  {(b)} The best-fit real-space electron density profile (solid lines) reveals a modulated water structure on graphene in agreement with the prediction of molecular dynamics (MD) simulations (dashed lines). {(c)} The intrinsic water density profiles next to a single uniform graphene surface from the experiment (red solid line) and from MD simulations (dashed lines; O: red, H: blue, C: gray); the insets show MD snapshots of the water-graphene interface along the surface normal direction (left) and in the plane within $z<0.5$ nm  (right); the arrows represent the water dipole moment orientation, and the dashed lines represent temporary hydrogen bonds. {(d)} Water orientation probability distribution function $\mathscr{P}(\cos \Omega)$ within different water regions: $h_1\equiv\{0~\leq z< 0.5 {\rm nm}\}$, $h_2\equiv\{0.5~{\rm nm}\leq z< 1~{\rm nm}\}$, and $h_3\equiv\{1~{\rm nm}\leq z< 1.5~{\rm nm}\}$ (see Figure \ref{fig0}a). The water orientation is given by $\cos \Omega$, where $\Omega$ is the angle between a water molecule dipole moment ({\boldmath${\upmu}$}) and a unitary vector normal to the graphene surface ({\bf e}$_{_\bot}$); $\upmu_0\equiv${\boldmath$|\upmu|$}=0.0489 $e\cdot$nm is the dipole moment of the SPC/E water model, where $e$ is the positive elementary charge; from left to right the three insets show representative configurations of water for $\cos(\Omega)=-1,0$, and $\cos(\Omega)\gtrsim$ 0, respectively. {(e)-(g)} The histograms of the instantaneous water polarization components  $H({\rm p}_{i})$ ($i= x,y,z$). P$_x$ and P$_y$ are always equal to zero. The polarization histograms in the $z$-direction are persistently narrower than in the $x$- and $y$-directions up to a water layer thickness of $L_{\rm w}\approx$ 120 nm.}
\label{fig2}
\end{figure*}

{Here, we focus on the details of the water structure in the absence of ions and the relationship between XR experiments and the simulation results}. Analysis of the XR data (red line in Figure \ref{fig2}a) reveals a graphene/water electron density profile with four partial surfaces to which water adsorbs (Figure \ref{fig2}b). \hl{A comparison with XR data of graphene in air (blue line in Figure 4a) reveals the sensitivity of the measurement to the water structure.} These include three layers of graphene, G$_1$, G$_2$, and G$_3$ with fractional layer coverages of 0.84 ML, 0.31 ML, and 0.1 ML, respectively, and a reconstructed carbon buffer layer, G$_0$ with complete coverage and which separates true 2D graphene (G$_1$-G$_3$) from the SiC substrate below (see the complete interface structure in Appendix C and Table \ref{Table:Fitparams} for details). 

In general, the XR measurements and MD simulations reveal qualitatively equivalent water distributions adsorbed on the partial graphene surfaces (Figure \ref{fig2}b). Both the XR-derived structure and the MD prediction show a weakly modulated water profile with density peaks that correlate with the locations of the graphene layers. The XR best-fit structure shows water adsorbed at $\approx0.31 \pm 0.03$ nm above each exposed graphene surface while MD predicts the adsorbed water height to be 0.33 nm, both of which are consistent with a slightly hydrophobic interface~\cite{Uysal_hyrophobic,Mezger_hydrophobic}. 

Zhou \textit{et al.} found, via XR measurements, a significantly reduced water height of 0.23 nm on the G$_0$ buffer layer compared to a height of 0.32 nm above free-standing graphene, which suggested that the G$_0$ layer exhibits a hydrophilic character. In contrast, their AIMD results showed a $\sim$0.02 nm decrease in the water height above G$_0$ ($z\sim$0.31 nm) compared to that above free-standing graphene layers ($z =$ 0.33 nm), which is in agreement with our XR results. Zhou \textit{et al.} attributed the discrepancy in G$_0$ water height between their XR and AIMD results to surface defects in their EG/SiC sample, which likely also explains the observations in the present study. We discuss sources of the G$_0$ water height discrepancy in greater detail in Appendix C. Finally, MD predicts a broadening of the first hydration layer for thicker graphene regions (i.e., on G$_3$ where the total thickness of the graphene slab is $\sim$1 nm ), a phenomenon not observed in the XR results, with the root mean square (RMS) width of water on G$_3$ being most consistent with the experimental results (see Appendix C). Potential explanations of the subtle differences between our MD prediction and XR results may reflect different interactions between water and graphene of various thicknesses and finite size effects of the simulation compared to the relatively large area of the XR measurement (see Appendix C).


The intrinsic water structure is the density profile corresponding to a uniform graphene surface (see Figure \ref{fig0}a) extracted from XR measurements. Figure \ref{fig2}c shows a comparison of the intrinsic water density profiles from XR experiments and MD simulations. Both the XR and MD oxygen density profiles reveal a first hydration layer ($z\approx$0.3 nm) with a peak density that is more than twice that of the bulk. The density oscillations decay rapidly with a small secondary hydration layer at $z\approx$0.6 nm and a nearly bulk-like third hydration layer at $z\approx$1 nm in both XR and MD profiles. The atomic positions of hydrogen are calculated via MD simulations only due to the weak scattering of X-rays from hydrogen atoms. \hl{As such, the agreement between experimental and computational oxygen distributions suggests a layered proton distribution (according to our MD simulations) leading to a non-zero interfacial water polarization.} The position of the oxygen layers relative to the graphene surface (validated by the experiment) are used to build the histograms of the interfacial water polarization which is employed to explain the electrostatic interfacial interactions  (see below).

The location of the hydrogen density peak at the first hydration layer coincides with the oxygen density peak but is broader, suggesting that a fraction of the water dipole moments are oriented perpendicularly away from the graphene surface. This is partially driven by transient hydrogen bonds between water molecules of the first and second hydration layers (see left inset in Figure \ref{fig2}c) and the inability of water molecules to form hydrogen bonds with the graphene surface. The water orientation is quantified by the distribution function of its dipole orientation $\cos \Omega$ with respect to the graphene surface  normal {\bf e}$_\bot$ (Figure \ref{fig2}d). The distribution reveals a first hydration layer in the region $h_1$ ($z<0.5$ nm) with the majority of water dipole moments oriented parallel to the graphene surface ($\cos \Omega = 0$). However, the asymmetry in the distribution exhibits a slight preference of the water dipole moment to orient away from the graphene surface.  The water molecules in the $h_1$ region form a  hydrogen bond network which is a characteristic behavior of water near hydrophobic molecules~\cite{acscentsci.8b00076} (right inset in Figure \ref{fig2}c). In the region defined at $0.5$ nm$<z\leq 1$ nm ($h_2$) the  preferential orientation of water molecules is diminished (Figure \ref{fig2}c), and from $z>1$nm all the water dipole moment orientations are equally probable ($1$ nm$<z\leq 1.5$ nm). 

\begin{figure*}
\includegraphics[trim={0.5in 0.1in 0.55in 0in},clip,width=12cm]{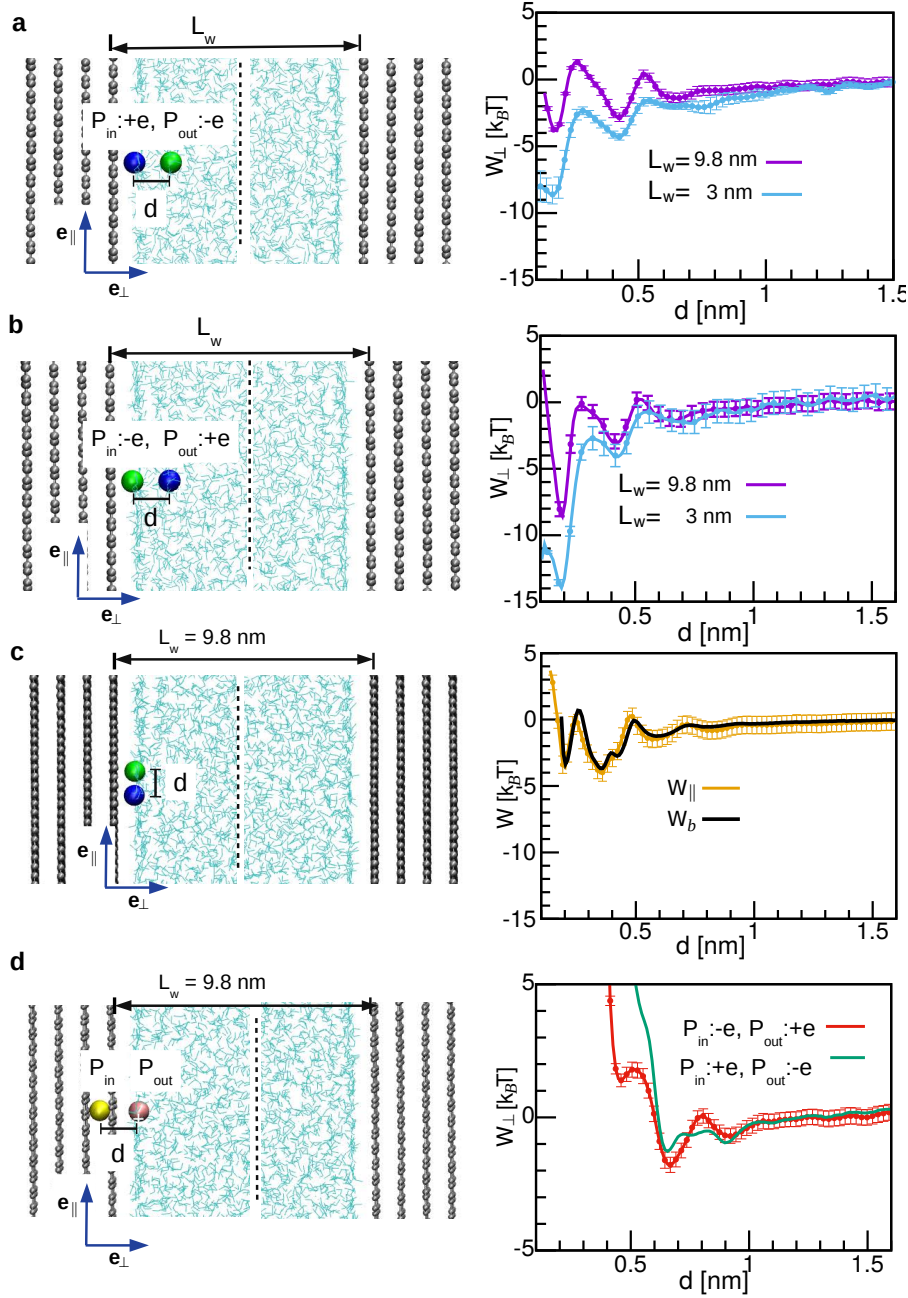} 
\caption{The ion-ion interaction under confinement is directional, non-equivalent upon permutation of the ions' position, and confinement-dependent ($L_{\rm w}$). The potential of mean force profiles along the graphene surface normal direction (${W}_{\bot}$) are investigated as a function of the ion-ion separation distance $d$. The  two ions (designated as P$_{\rm in}$ and P$_{\rm out}$) are placed at different heights $z$ above the graphene surface and the two ions have the same $x$ and $y$ coordinates; the P$_{\rm in}$ ion is fixed at $z_{\rm in} \approx$ 0.28~nm. The ion positions are exchanged such that in {(a)} P$_{\rm in}=$+e and P$_{\rm out}=$-e while in {(b)} P$_{\rm in}=$-e and P$_{\rm out}=$+e. The right hand side panels show the anion-cation potential of mean force at two confining separation distances, $L_{\rm w}=3$ nm (light-blue line) and  9.8 nm (purple line). The attractive interactions are non-reciprocal upon exchange the ion's positions and are enhanced by decreasing $L_{\rm w}$. {(c)} The in-plane potential of mean force profile (${W}_{||}$) between two ions fixed at $z \approx$ 0.28 nm and with variable separation distance $d$ ($L_{\rm w}=9.8$ nm, yellow line) is similar to that in the bulk ${W}_{b}$ (black line). {(d)} The intercalation of one of the ions between the G$_2$ and G$_3$ graphene layers leads to predominantly repulsive forces. For the $W_{\bot}$ calculations $d$ is varied along the surface normal direction, while $d$ is varied along the surface parallel direction for the $W_{||}$ calculations; in the bulk $W_b$ is isotropic.} 
\label{Fig3}
\end{figure*}


{We now examine in further detail the interfacial water polarization. In recent experiments it is investigated the out-of-plane permittivity ($\varepsilon_\bot$) of water confined  between two flat surfaces of graphite and hexagonal boron nitride. The experiments reveal that $\varepsilon_\bot \sim$ 2 within an interfacial layer of two or three molecules thick (referred to as the electrically dead layer) which is considered to have a ``vanishingly small polarization"} \cite{Fumagalli1339}. {Here in contrast, we observe a non-zero persistent water polarization in the $z$-direction within the interfacial region.  Figures} \ref{fig2}e,f,g show  the histograms of the instantaneous polarization  $H(\rm p)$ within the regions $h_1$, $h_2$, and $h_3$, respectively, defined to capture the different hydration layers. The average polarization in the $z$-direction is P$_z\equiv\langle$p$_z\rangle\approx 0.03 e/{\rm nm}^{2}$ in the $h_1$-region (Figure \ref{fig2}e), P$_z\approx$-0.004$e/$nm$^{2}$ in the $h_2$-region (Figure \ref{fig2}f), and P$_z$=0 in the $h_3$-region (Figure \ref{fig2}g) and for $z\gtrsim$ 1 nm, while P$_x$ and P$_y$ are always equal to zero. {In particular we highlight the non-vanishing polarization $P_z$ within the $h_1$ region which is fundamental to understand the electrostatic interactions near an interface (see next section).}

The dipole fluctuations are known to be suppressed along the surface normal direction \cite{jz401108n}. Here, the suppression is seen as a much narrower histogram for the polarization in the $z$-direction than in the $x$- and $y$-directions. The  polarization suppression implies that every fluctuation of $\upmu_{z}$ is nearly canceled by an antiparallel component. The suppression is observed  across the entire water region and up to a water layer thickness of $L_{\rm w}\approx$ 120 nm. Additionally, we investigated the liquid-vapor interface by removing the graphene surfaces, and we found that the dipole moment fluctuations in the $z$-direction remain suppressed. \hl{This observation shows that the suppression is not only observed at a liquid-solid interface, and it is not related to the chemical structure of the confining surfaces.}

%

\begin{figure*}[h]
\includegraphics[trim={0.2in 0.1in 0in 0in},clip,width=12cm]{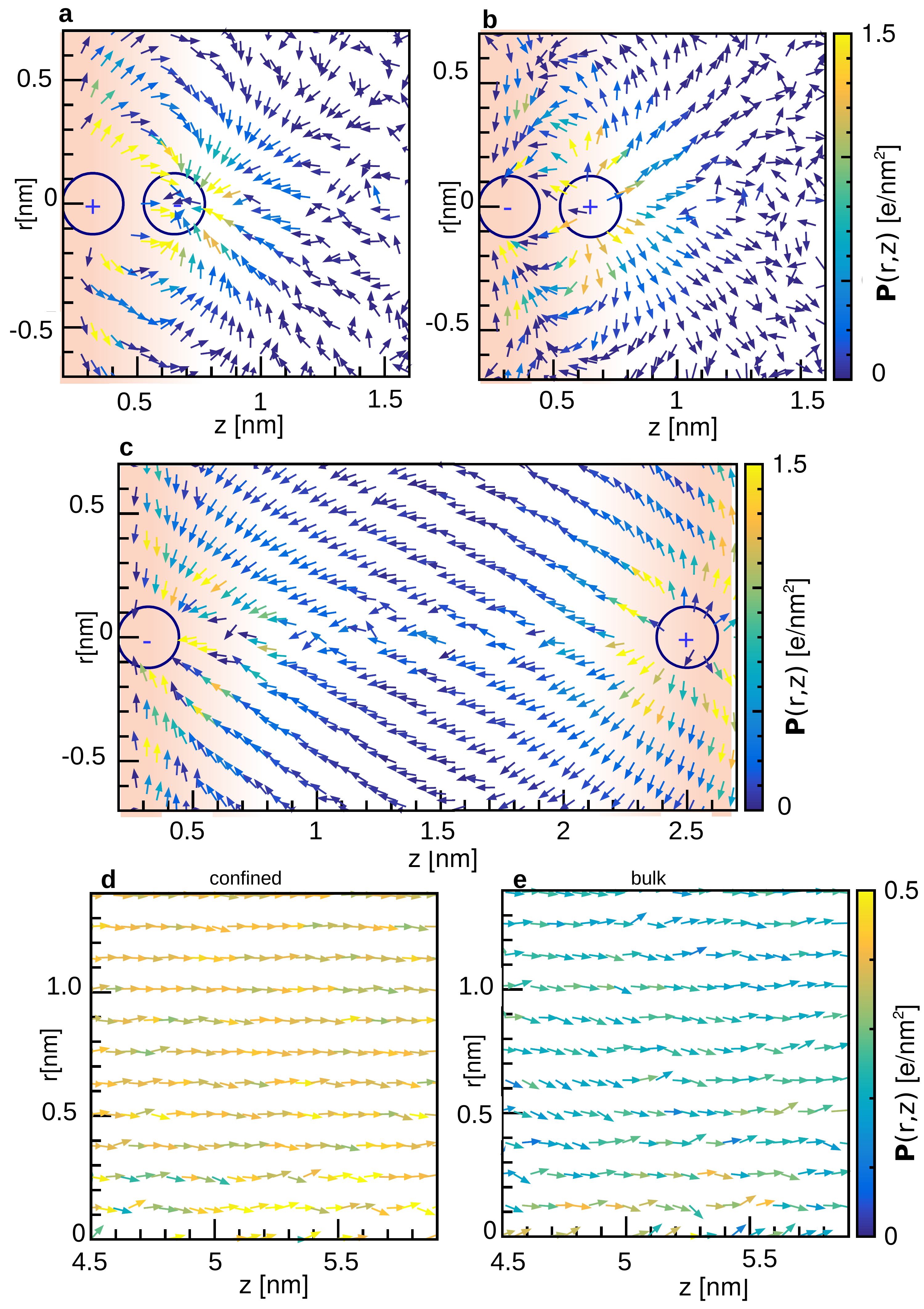}
\caption{Water polarization field in confinement affects the ion-ion effective interaction. The anion-cation separation vector is oriented along the graphene surface normal; {(a)} the cation is next to the graphene surface  at $z_{\rm in} \approx$ 0.28 nm and the anion is at $z_{\rm out} \approx$ 0.62 nm; {(b)} the anion is  at $z_{\rm in} \approx$ 0.28 nm while the cation is at $z_{\rm out} \approx$ 0.62 nm. {(c)} The anion and cation are placed at a separation distance of 2.2 nm  at the opposite sides of the water layer near the graphene surfaces separated by a water layer of thickness $L_{\rm w}=3$~nm. The water polarization at the middle region between the ions at a separation distance of 8.9~nm in {(d)} confinement ($L_{\rm w}=9.8$~nm) and in {(e)} bulk. The bulk system is simulated using periodic boundary conditions in the $x$, $y$, and $z$ directions in a box of length $L_z=20$ nm in the $z$ direction.} 
\label{Fig4}
\end{figure*}

\clearpage
\subsection{Ion-Ion Interactions in Confinement} 
We look into a direct force analysis between an anion and a cation in bulk water, near the graphene surface, and between an ion in the water and an ion intercalated between the graphene layers. Ion intercalation is the common mechanism for Faradaic energy storage (e.g., lithium ion batteries with graphitic anodes) \cite{ji2019lithium}, while ion adsorption occurs in capacitive energy storage systems. Figures \ref{Fig3}a-c show that confinement enhances the ion-ion interaction along the surface normal direction, which is more pronounced at $L_{\rm w}=$ 9.8 nm than in the bulk and even more pronounced by decreasing the water layer thickness to $L_{\rm w}$=3 nm; the minimum of the PMF decreases by about $5k_BT$ by decreasing $L_{\rm w}$ from 9.8 nm to 3 nm. The enhancement of the ion-ion interaction near the graphene surface is consistent with observations near a hydrophobic surface~\cite{Chen555}. Furthermore, the effective ion-ion interaction  under confinement is non-reciprocal, i.e., the potential of mean force is not equivalent upon exchange of the ions' positions (See Figure \ref{Fig3}a, b) where it is seen that the minimum of the PMF decreases by about $5k_BT$ by exchanging the ions' positions. {This non-reciprocal behavior of the anion-cation interactions is also observed when polarizable models of Na$^+$ Cl$^-$ ions, polarizable water, and polarizable graphene are used in the simulations (see Appendix} \ref{sec.polarizability}). The ion-ion interaction in the surface plane direction $W_{||}$, however, is not significantly different from the ion-ion interaction in bulk (Figure \ref{Fig3}c) likely because the ions are less confined in the $xy$-plane. Interestingly, the interaction along the surface normal direction changes from predominantly attractive when both ions are in the aqueous phase  (Figure \ref{Fig3}a,b) to completely repulsive when one ion is intercalated within the graphene interlayer space (Figure \ref{Fig3}d).  In an isotropic medium the ion-ion effective force depends only on the ions' separation distance, and the forces are equivalent upon exchange of the ions' positions. However, figures \ref{Fig3}a-d reveal that in confinement the effective ion-ion interaction is directional and position dependent with respect to the graphene surface. The force non-reciprocity cannot be explained by current models of anisotropic dielectric permittivity ($\varepsilon_{||}$ and $\varepsilon_{\bot}$), which predict the same dielectric profile regardless of the ion valence and, hence, a symmetric force. We will discuss this point in detail in subsection \ref{sec:res_molect}.

The non-reciprocal behavior of the ion-ion interfacial interactions and the ion-sign specificity are due to the unsymmetrical response of the interfacial water polarization. The classical theory assumes surface polarization is the same for positive and negative charges whereas our results show the polarization field strongly depends on the sign of the charge. The potential of mean force $W_c$ from the MD simulations takes into account the ion-ion direct interaction and the interaction through the surrounding water molecules, which includes the work to reorient the interfacial polarization.  The interfacial polarization when the cation is placed on the graphene surface and when the anion is on the surface (exchanged configuration) are shown in Figures \ref{Fig4}a and \ref{Fig4}b, respectively.  These polarization fields give rise to different ion-ion interactions. The polarization around two ions close to each other is canceled at a short distance ($\lesssim$ 1.5 nm) which is consistent with the observation around the polar groups of proteins \cite{Qiao19274}. When the two charges are separated at longer distances, however, the polarization propagates over the whole separation distance (Figure \ref{Fig4}c).  The polarization in confinement is enhanced even at larger confinement distances. Figures \ref{Fig4}d,e show the difference in the water polarization at the middle distance between the two ions separated by about 9.5 nm in confinement and in bulk, respectively. The magnitude of polarization in confinement is approximately twice and shows less deviation from the perpendicular orientation than the polarization in bulk.

\begin{figure*}[ht]
    \includegraphics[trim={0in 0in 0.0in 0in},clip,width=16cm]{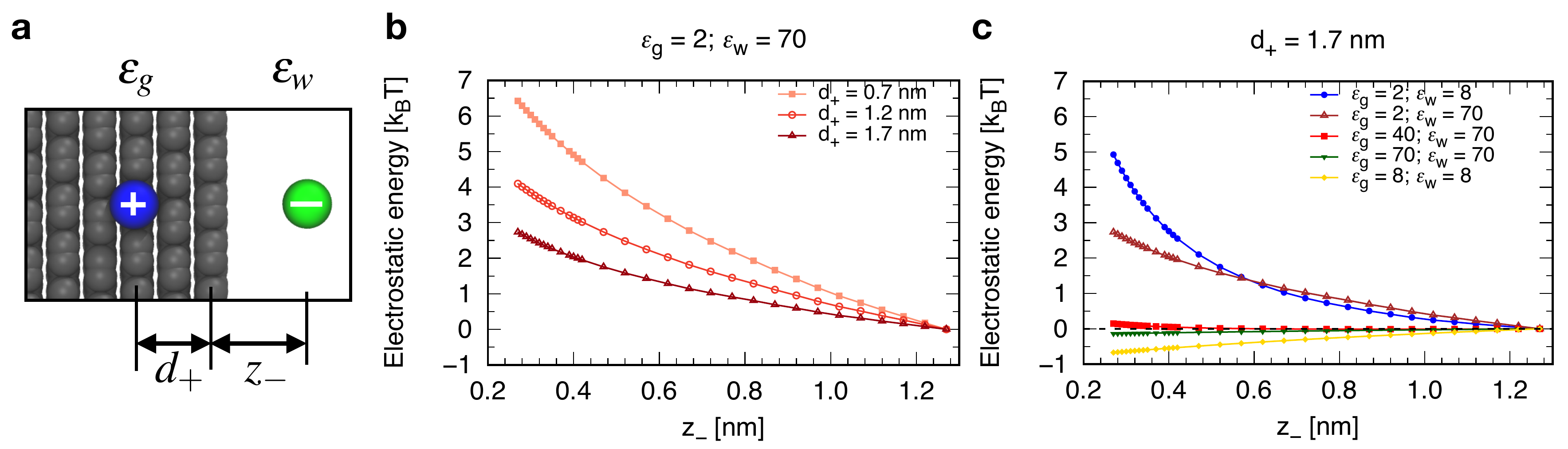}
    \caption{Coulombic energy as a function of the distance between the two ions computed from the image charge method for the case where the cation is inside the graphene surface. In (a) graphene atoms are visualized to guide the eye but not involved in the image charge calculations.}
    \label{fig:ecoul2}
\end{figure*}
\subsection{Results from continuum electrostatics theories}
\label{sec:res_continuum}

We calculate the ion-ion Coulombic interaction energy as a function of the distance between the two ions arranged in the three configurations in Figures \ref{Fig3}a-c using the image charge method (see subsection \ref{ssec:conti}). We assume that the water is a continuum background with the permittivity of $\varepsilon_{\rm 1} =\varepsilon_{\rm w} = 70$ and the graphene is a continuum medium of permittivity of $\varepsilon_{\rm 2} =\varepsilon_{\rm g} = 2$.  The image charge method cannot predict the oscillations found in the potential of mean force in Figure \ref{Fig3} when surface polarization effects are included. Instead, a purely repulsive interaction is observed assuming a dielectric mismatch at the interface of $\alpha= 0.97$. Figure \ref{fig:ecoul2} shows the ion-ion interaction energy (from the image charge method) when the cation is placed inside the graphene surface for different combinations of the dielectric constants of the two media forming the interface. It is interesting to see that the continuum theory captures the change from attractive to repulsive for sufficiently large dielectric mismatches. The repulsive behavior is in agreement with the results from umbrella sampling simulations when one ion is intercalated between the graphene layers (see Figure \ref{Fig3}d), however, the interaction energy obtained from the image charge method is the same when the ions' positions are exchanged.

\begin{figure*}
\includegraphics[trim={1.2in 0.1in 1.2in 0in},width=11cm]{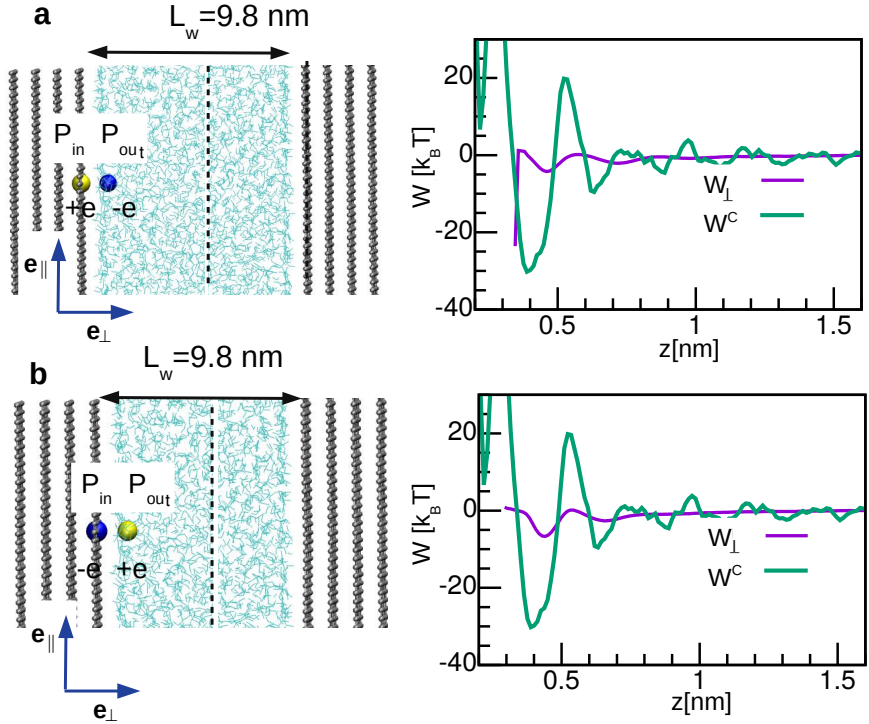}
\caption{Interaction between two ions at the water graphene interface along the surface perpendicular direction. The ion at the inner position P$_{\rm in}$  is placed as part of the outermost graphene layer while the ion at the outer position P$_{\rm out}$ is into the water phase, in {(a)} P$_{\rm in}=$+e and P$_{\rm out}=$-e while in {(b)} P$_{\rm in}=$-e and P$_{\rm out}=$+e. The right hand side panels show the anion-cation potential of mean force where the purple line (W$_\bot$) represents the potential of mean force calculated via umbrella sampling and the green line is the the interaction potential calculated as W$^{\rm C}=l_0\varepsilon^{-1}_{\bot}(z)/z$, where $l_0={e^2}/{(4\pi\varepsilon_0 k_B T)}$ and the inverse perpendicular permittivity $\varepsilon^{-1}_{\bot}(z)$ is calculated via Eq. \eqref{eqn1}.}
\label{f_cou}
\end{figure*}

\subsection{Results from the molecular theory of permittivity}
\label{sec:res_molect}

We investigated the interfacial dielectric permittivity employing the molecular theory described in subsection \ref{ssec:molect}. The expressions of the permittivity in the direction parallel to the graphene surface ($\varepsilon_{||}(z)$) and the inverse permittivity in the perpendicular direction ($\varepsilon^{-1}_{\bot}(z)$) are given by  Eqs. \eqref{eqn1} and \eqref{eqn2}, respectively. The ultimate goal of quantifying the interfacial permittivity is to determine the electrostatic interaction between charged species near the interface. We estimate the interfacial ion-ion interaction by means of the Coulomb's potential employing a position dependent permittivity  $\varepsilon({\bf r})$ given by Eq. \eqref{eqn2}. We see that the interaction potential W$^{\rm C}$ significantly overestimates the interaction with respect to the potential of mean force from umbrella sampling simulations W$_\bot$ (see Figure \ref{f_cou}). 
As we mentioned above, the potential of mean force from umbrella sampling simulations W$_\bot$ is different by exchanging the positions of the two particles (non-reciprocal) whereas here we see that the ion-ion interaction calculated from the permittivity profile from Eq. \eqref{eqn2} is the same independently of the ions arrangement (see Figure \ref{f_cou}). This shows that the interaction between two charged particles can not be described solely in terms of a permittivity function.

\section{Conclusions}

Our results show that in confinement the ion-ion effective interaction is directional and position dependent with respect to the graphene surface due to a non-zero persistent interfacial water polarization. We identify this property as non-reciprocity, which describes the directional dependent interactions and non-equivalent change of interactions upon exchange of the ions' positions near a confining surface. The non-reciprocity implies that ion-ion interactions at the interface do not obey the isotropic and translational symmetries of Coulomb's law {and are observed in both polarizable and non-polarizable models.} This phenomenon contrasts with the ion-ion interactions in an isotropic medium (bulk) where the force depends only on the ions' size and separation distance, is not directional, and is equivalent by exchange of the ions' positions. 
Traditionally, ion specificity is attributed to the internal ion polarization including polarizability associated with ion size. Here, we find that the water polarization plays a central role in the behavior of ions near interfaces. Namely, the water polarization around ions and near the interface alters electrostatic interactions, leading to non-equivalent ion-interface interactions upon exchange of the ion charge sign even if the ions have equal size. This non-symmetrical water polarization affects the understanding of ion-differentiation mechanisms such as ion selectivity and ion specificity. \hl{The agreement between XR experimental measurements and MD simulations of polarizable and non-polarizable models suggests a layered hydrogen structure which leads to the interfacial water polarization. We find that the water structure near the graphene-water interface, however, is not enough to infer electrostatic interactions near the interface. The current models based on the anisotropic dielectric permittivity in confinement (obtained via the water structure) cannot explain the non-reciprocal ion-ion interactions found here. Our simulations reveal that the valence (charge) of the ions, irrespectively of other ion specific effects, is responsible for the non-reciprocal interactions.} Molecular interactions near interfaces and in confinement are related to a variety of processes including chemical reactions, adsorption, and biological molecular recognition. The insights gained here need to be considered in the understanding of processes based on asymmetric ionic adsorption and interactions at heterogeneous interfaces such as proteins.


\newpage
\section*{Acknowledgments}
F.J.A., T.D.N. and M.O.d.l.C. were supported by the Department of Energy (DOE), Office of Basic Energy Sciences under Contract DE-FG02-08ER46539. The computational work was done with the support of the Sherman Fairchild Foundation. This work was supported, in part, by the Midwest Integrated Center for Computational Materials (MICCoM) as part of the Computational Materials Sciences Program funded by the U.S. Department of Energy, Office of Science, Basic Energy Sciences, Materials Sciences and Engineering Division (5J-30161-0010A). 
K.J.H. gratefully acknowledges support from the U.S. Department of Defense through the National Defense Science and Engineering Graduate Fellowship (NDSEG) Program and from the Ryan Fellowship at Northwestern University International Institute of Nanotechnology. We thank Dr. Jon Emery (Northwestern University) and Dr. D. Kurt Gaskill (Naval Research Laboratory) for the graphene sample and for discussions on the structure of graphene. X-ray reflectivity measurements were performed at beamline 33-ID-D of the Advanced Photon Source at Argonne National Laboratory (ANL), a U.S. DOE Office of Science User Facility operated by ANL under Contract No. DE-AC02-06CH11357. 

\clearpage

\appendix 
\section{{Polarizable Models}}
\label{sec.polarizability}

\begin{figure}[!h]
\includegraphics[trim={1.2in 1.2in 1.2in 1.2in},width=7.5cm]{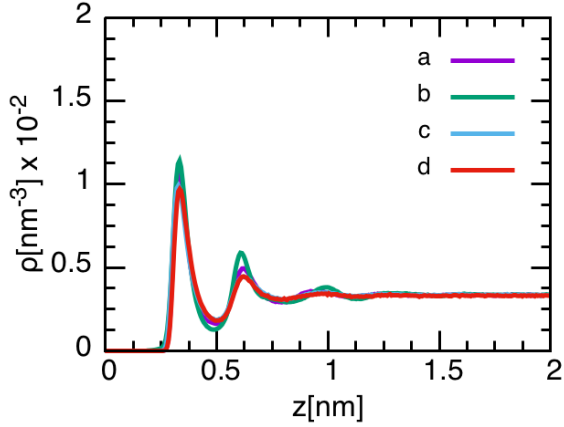}
\caption{{Effect of water and graphene polarizabilities on the water density profile. (a) Polarizable water ($\alpha_{\rm w}=0.978$ \AA$^3$) and polarizable graphene ($\alpha_{\rm g}=1.139$ \AA$^3$), (b) polarizable water ($\alpha_{\rm w}=0.978$ \AA$^3$) and non-polarizable graphene ($\alpha_{\rm g}=0$), (c) non-polarizable water ($\alpha_{\rm w}=0$) and polarizable graphene ($\alpha_{\rm g}=1.139$ \AA$^3$), (d) non-polarizable water ($\alpha_{\rm w}=0$) and non-polarizable graphene ($\alpha_{\rm g}=0$). The water layer thickness is $L_{\rm w}=9.8$ nm.}}
\label{fig11}
\end{figure}

{The induced polarization of atoms results from the deformation of the electronic cloud due to a local electric field and it is known to be relevant in some interfacial phenomena}~\cite{la204036e}. {At first approximation, the main contribution of the induced polarization is from the induced dipole moment ({\boldmath$\upmu$}$_{\rm ind}$) of atoms, ions and molecules due to the local electric field {\bf E}'$({\bf r})$, given by}

\begin{equation}
 {\boldmath \upmu}_{\rm ind} = \alpha_i  {\bf E}'({\bf r})
\end{equation} 
{where $\alpha_i$ is the polarizability, which is a distinctive property for each atom and ion. To account for polarization effects, we employ the SWM4-NDP polarizable water model} \cite{LAMOUREUX2006245} {and the corresponding force field parameters for the polarizable Na$^+$ and Cl$^-$ ions} \cite{ct900576a}. {Polarizable graphene is modeled by including the polarizability derived from DFT calculations }\cite{jpcc.7b08891} {in our graphene model. The water molecule has a fixed HOH geometry bearing two positive charges at the hydrogen centers and the negative charge is placed at a fixed distance from the oxygen center, shifted along the molecule's axis of symmetry. The induced polarization is based on the Drude particle model of charge $q_D$, attached to the center of the polarizable atoms (or ions) through a harmonic potential. The Drude particle's charge is balanced by the positive charge of the core such that, $q_C+q_D=0$ for neutral atoms and $q_C+q_D=q_{\rm ion}$ for ions, where $q_{\rm ion}$ is the ionic charge. The spring constant $k_D$, the polarizability $\alpha_i$, and the charge of the Drude particle $q_D$, are related by}

\begin{equation}
\alpha_i = q^2_D/k_D 
\end{equation}
{The polarizabilities employed in our calculations are $\alpha_{\rm g}=1.139$ \AA$^3$ for graphene carbon atoms, $\alpha_{\rm w}=0.978$ \AA$^3$ for water oxygen atoms, $\alpha_{\rm Na^+}=0.157$ \AA$^3$ for Na$^+$ and,  $\alpha_{\rm Cl^-}=3.969$ \AA$^3$  Cl$^-$.}
{Integration of the equations of motion is performed by means of the extended Lagrangian algorithm} \cite{1.1589749} {which consists in assigning a small mass to the Drude particles $m_D$ taken from the atomic masses, in such a way that the mass of the core is $m_i-m_D$. The Drude particles are simulated at a much smaller temperature than the whole system to fulfill the Born-Oppenheimer minimum energy condition. The Drude particles mass is $m_D=0.4$ g/mol. A dual Nos\'e-Hoover thermostat is used to maintain the Drude particles temperature at $T_D=$ 1K and the system at $T=298$ K. We simulated a set of systems similar to that described in subsection} \ref{ssec.models}, {namely, a water layer of thickness $L_{\rm w}\approx$ 9.8 nm, formed by $N_{\rm w}=8060$ water molecules, and confined between two graphene surfaces formed by four graphene layers each} (see Figure \ref{fig0}a). {A time step of 1 fs is employed for integration of the equations of motion of the systems without ions. When ions are present the time step is 0.2 fs. The simulation protocols are similar to those described in Section} \ref{sec.methods}. {Consideration of polarizability of the water molecules implies an additional dipole moment contribution to the permanent dipole moment.}

\begin{figure}[!h]
\includegraphics[trim={1.2in 0in 1.2in 0in},width=7.5cm]{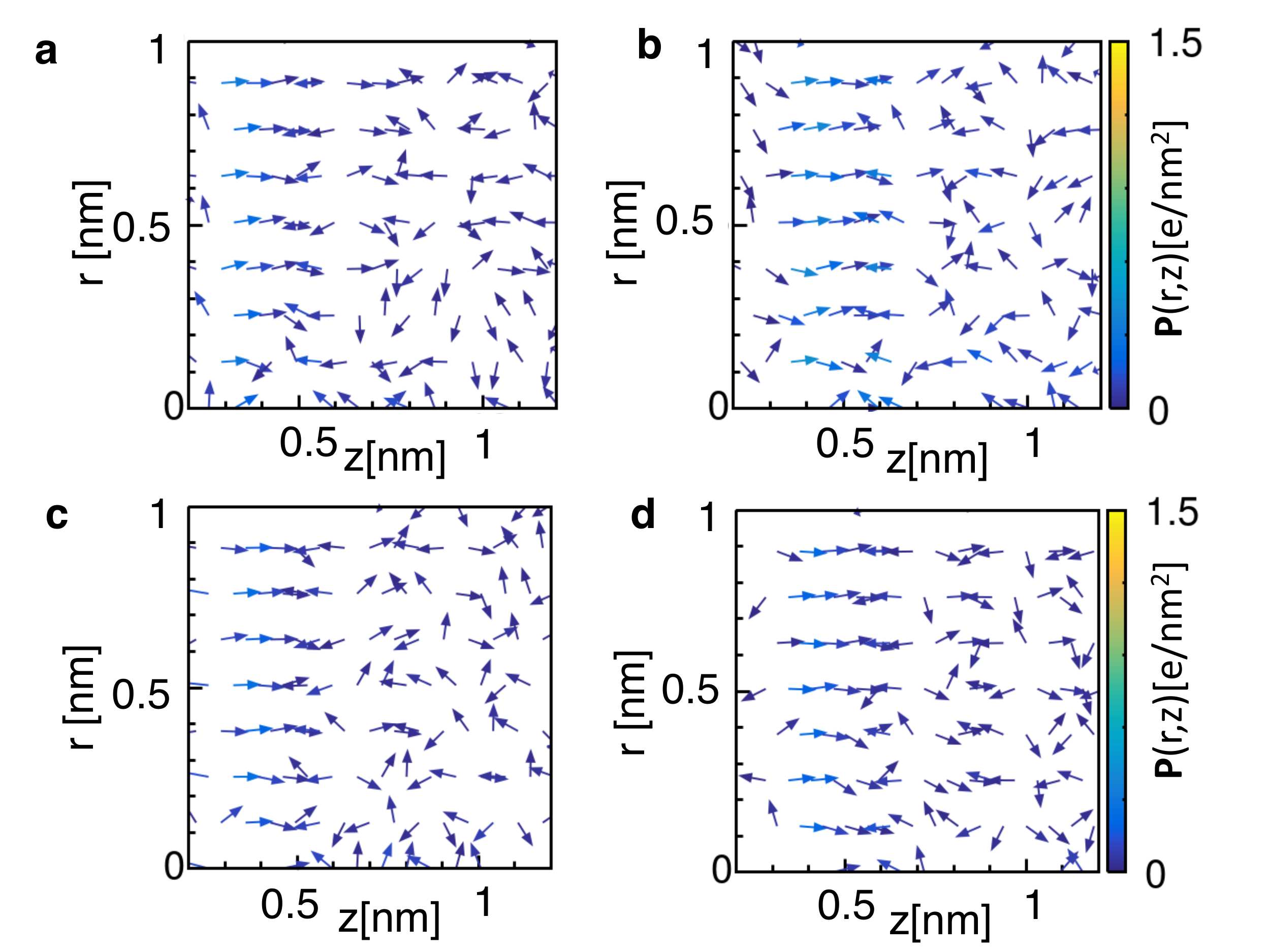}
\caption{{Polarization field in polarizable models. (a) Polarizable water ($\alpha_{\rm w}=0.978$ \AA$^3$) and polarizable graphene ($\alpha_{\rm g}=1.139$ \AA$^3$), (b) polarizable water ($\alpha_{\rm w}=0.978$ \AA$^3$) and non-polarizable graphene ($\alpha_{\rm g}=0$), (c) non-polarizable water ($\alpha_{\rm w}=0$) and polarizable graphene ($\alpha_{\rm g}=1.139$ \AA$^3$), (d) non-polarizable water ($\alpha_{\rm w}=0$) and non-polarizable graphene ($\alpha_{\rm g}=0$). }}
\label{fig12}
\end{figure}

{First we examine the water structure near the graphene surface considering four different combinations, namely, a) polarizable water and polarizable graphene,  b) polarizable water and non-polarizable graphene, c) non-polarizable water and polarizable graphene, and d) a) non-polarizable water and non-polarizable graphene. The density profiles as a function of the perpendicular distance to the graphene surface are shown in Figure} \ref{fig11}. {The four combinations predict qualitatively similar profiles exhibiting peaks at the same location, approximately. Our results are consistent with previous studies which show that the graphene polarizability does not significantly affect the water density profile}~\cite{jpcc.7b08891,1.4789583}. {Here, the polarizable water model predicts higher peaks the non-polarizable water. Analysis of the polarization field shows a persistent interfacial polarization qualitatively similar in the four combinations }(see Figure \ref{fig12}).

\begin{figure}
\includegraphics[trim={1.2in 1.2in 1.2in 1.2in},width=8cm]{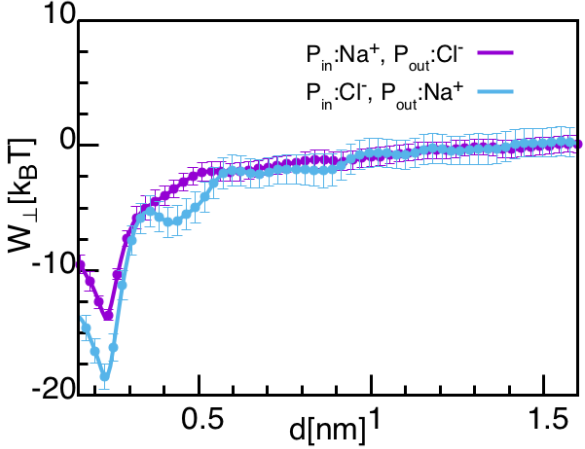}
\caption{{Interaction between polarizable Na$+$ and Cl$^-$ ions at the water-graphene interface. The potential of mean force (${W}_{\bot}$) as a function of the ion-ion separation distance $d$ along the graphene surface normal direction. The  two ions (P$_{\rm in}$ and P$_{\rm out}$) are placed at different heights $z$ above the graphene surface at the same $x$ and $y$ coordinates (see Figure} \ref{Fig3}a,b). {The ion positions are exchanged such that in {(a)} P$_{\rm in}=$Na$+$ and P$_{\rm out}=$Cl$^-$ while in {(b)} P$_{\rm in}=$Cl$^-$ and P$_{\rm out}=$Na$^+$. The ions' polarizabilities are $\alpha_{\rm Na^+}=0.157$ \AA$^3$ and $\alpha_{\rm Cl^-}=3.969$ \AA$^3$ for Na$^+$ and Cl$^-$, respectively. The water polarizability is $\alpha_{\rm w}=0.978$\AA$^3$ and the graphene polarizability is $\alpha_{\rm g}=1.139$\AA$^3$. The water layer is confined between two graphene surfaces separated by $L_{\rm w}=9.8$ nm.}}
\label{fig13}
\end{figure}

{In Figure} \ref{fig13} {we analyze the effective interaction between two ions (Na$^+$ and Cl$^-$) in water nearby the graphene surface.  All the components in the system (water, ions, and graphene) are polarizable. The ions are placed at different heights above the graphene surface aligned perpendicularly to the graphene surface in a similar way as in Figures} \ref{Fig3}a,b. {The potential of mean force (PMF) is calculated in two cases: 1) the Na$^+$ ion is placed on the graphene surface while the Cl$^-$ ion is at different heights from the graphene surface and 2) the ions' positions are exchanged. The PMF profiles are qualitatively different from those for-non polarizable ions (see Figure} \ref{Fig3}a,b {for $L_{\rm w}=9.8$ nm). For example, the profiles exhibit less pronounced oscillations when Na$^+$ is on the graphene surface (purple line in Figure }\ref{fig13}) {than for the non-polarizable ions discussed Figures }\ref{Fig3}a,b. {The difference in energy when the ion positions are exchanged, at the minimum of the PMF, is 5 $k_BT$ in the fully polarizable system (see Figure }\ref{fig13}). {This is similar in both magnitude and sign to that observed for the non-polarizable system (see Figure }\ref{Fig3}a,b {for $L_{\rm w}=9.8$ nm). Hence, the non-reciprocal behavior of the ion-ion interactions is present in both systems with polarizable and non-polarizable atoms and ions.}

\section{Linear Response Theory for the Dielectric Response}
\label{sec.permittivity}

An electric field ${\bf E}$ applied in a dielectric material induces a polarization due to a separation of the bound charges in the material (atomic nuclei and their electrons). The macroscopic field ${\bf D}$ is called the electric displacement and is given as
\begin{equation}
 {\bf D}= \varepsilon_0 {\bf E}+{\bf P} 
 \label{disp_1}
\end{equation}
where ${\bf P}$ is the polarization density. Linear response theory assumes that for a weak applied electric field the, induced electric polarization is proportional to the magnitude of the applied field 

\begin{equation}
{\bf P} =\varepsilon_0 {\underline{\chi}}\cdot {\bf E}
\end{equation}
where ${\underline{\chi}}$ is the {susceptibility} tensor, which is related to the dielectric permittivity tensor ${\underline{\varepsilon}}({\bf r})$ by
\begin{equation}
 {\underline{\varepsilon}} = {\bf I} + {\underline{\chi}}.
 \label{tensor}
\end{equation}
{{\bf I} is the identity matrix. {\bf D} is then} expressed as 
\begin{equation}
 {\bf D}=  \varepsilon_0 {\underline{\varepsilon}}\cdot {\bf E} 
 \label{disp_2}
\end{equation}
Using index notation { in} Eq. \eqref{disp_2} in Cartesian coordinates can be expressed as
\begin{equation}
 { D}_{\alpha}=  \varepsilon_0\sum_{\beta=x,y,z}{{\varepsilon}_{\alpha \beta}}{E_{\beta}}. 
\end{equation}
In simple materials the dielectric behavior is isotropic{, leading to a diagonal} dielectric tensor {with three equal} components. { Under nanoscale confinement} the dielectric response is anisotropic, and the components of the dielectric tensor are not equal. In the next part we present the main steps to derive the expression for the dielectric permittivity for a non-uniform medium in one dimension. 


Our treatment follows closely the work by Feller and Stern \cite{feller_JCP2003}, Ballenegger and Hansen \cite{Hansen_JCP2005}, and Roland Netz' group \cite{Netz_la2012}. The basic idea is to compute the response of a dielectric medium to a static, external, and uniform electric field {\bf E}, combining descriptions from statistical mechanics and continuum electrostatics. By combining both approaches, we are able to derive the expression of the local permittivity. 

Let $\Delta {\bf E}({\bf r})$  be the change in the mean local electric field inside the dielectric{. W}hen the external electric field  {${\bf E}$} is {turned} on, the change is due to {both} the external field itself and the dipoles {within} the medium.  {${\bf E}_0({\bf r})$} and ${\bf P}_0({\bf r})$ are, respectively, the mean electric field and the mean local polarization with no applied external field (${\bf E}_0({\bf r})$ is zero if ${\bf P}_0({\bf r})$). In the linear response regime the change in the local polarization {$\Delta{\bf P}({\bf r})$} and the change in the total electric field {$\Delta {\bf E}({\bf r})$} are related by

\begin{equation}
\Delta {\bf P}({\bf r}) ={\underline{\chi}}({\bf r})\cdot {\Delta \bf E}({\bf r})
\label{pol.7}
\end{equation}
where ${\underline{\chi}}({\bf r})$ is the local susceptibility tensor which is related to the local dielectric permittivity tensor ${\underline{\varepsilon}}({\bf r})$ as
\begin{equation}
 {\underline{\varepsilon}}({\bf r})={\underline{\chi}}({\bf r})+ {\bf I}
 \label{tensor2}
\end{equation}
The expression for $\Delta {\bf P}({\bf r})$ is derived from statistical mechanics while the expression for $\Delta {\bf E}({\bf r})$ is from macroscopic electrostatics. 

\subsection{Microscopic description}

We consider a classical system in a microstate $\Gamma$ described by the Hamiltonian $H(\Gamma)$. In general, the dipole moment {\bf m} changes from point to point within the dielectric{. H}ence, the instantaneous polarization density   at ${\bf r}$, ${\bf p}({\bf r})$, (also known as electric polarization, or simply polarization) is given by

\begin{equation}
 {\bf p}({\bf r}) = \frac{\Delta {\bf m}}{\Delta V}
\end{equation}

The system's total dipole ${\bf M}$ is given by

\begin{equation}
{\bf M} = \int {\bf p}({\bf r}) {\rm d}V
\end{equation}

The mean polarization is expressed as 

\begin{equation}
{\bf P}({\bf r}) = \langle {\bf p}({\bf r}) \rangle = \frac{\int{\bf p}({\bf r}) e^{-\beta H}{\rm d}\Gamma}{\int e^{-\beta H}{\rm d}\Gamma}
\end{equation}
Let $H$ and $H'$ be, respectively, the system's Hamiltonian when there is no applied external electric field and when the electric field is turned on. In linear response theory $H'$ can be expressed as $H'=H-{\bf M}\cdot{\bf E}$, and the change in the mean polarization as
\begin{eqnarray}
\Delta {\bf P}({\bf r}) &=& {\bf P}_{E}({\bf r}) - {\bf P}({\bf r}) \nonumber \\ 
&=& \langle {\bf p}({\bf r})\rangle_{E} - \langle {\bf p}({\bf r}) \rangle \\
&=&\frac{\int({\bf p}({\bf r}) - \langle {\bf p}({\bf r}) \rangle  ) e^{-\beta H'}{\rm d}\Gamma}{\int e^{-\beta H'}{\rm d}\Gamma} \nonumber
\end{eqnarray}
In the weak electric field regime we linearize the above expression {to} get
\begin{equation}
\Delta {P}_{\alpha}({\bf r}) = \beta \sum_{\gamma=x,y,z} [ \langle p_\alpha({\bf r}) M_{\gamma} \rangle -\langle p_\alpha({\bf r}) \rangle \langle M_{\gamma} \rangle]E_\gamma
\label{eq:1}
\end{equation}
where we have switched to express the components $\alpha=x,y,z$ of the vector $\Delta {\bf P}({\bf r})${,} and the statistical average is performed in the zero electric field regime. Eq. \ref{eq:1} involves the correlation between a fluctuation in the
local polarization density ${\bf m}({\bf r})$ and a fluctuation in the global dipole moment ${\bf M}$ of the system. 

To calculate the permittivity tensor via Eq. \eqref{pol.7} it is necessary to derive an expression for $\Delta {\bf E}$. The calculation of $\Delta {\bf E}$ is performed by considering the dipolar contributions from every molecule and from each image cell in a system where periodic boundary conditions are assumed in {the} $x$, $y$, and $z$ directions. By doing so, Feller and Stern arrived {at} the following expressions \cite{feller_JCP2003} for the dielectric permittivity profile in {the} parallel ($\varepsilon_{||}(z)$) and perpendicular ($\varepsilon_{\bot}(z)$) directions to the graphene surface given by Eqs. \eqref{eqn1} and \eqref{eqn2}, respectively.

\section{Experimental details}
\subsection{X-ray Reflectivity Background and Model}
The specular XR signal $R(Q)$ is directly related to the laterally-averaged real-space electron density profile $\rho(z)$ via Fourier Transform as
\begin{equation}
\label{eq:xrrefl}
R(Q) = T(Q)B(Q)\left(\frac{4\pi r_e}{A_{UC}Q}\right)^2\left|\int_{-\infty}^{\infty}{\rho(z)e^{iQz}dz}\right|^2    
\end{equation}
where $T(Q)$ is the angle-dependent transmission of X-rays through the sample cell, $B(Q)$ accounts for the effects of surface roughness \cite{roughness}, $r_e = 2.82 \times 10^{-5}$ {\AA} is the classical electron radius, and $A_{UC}$ is the unit cell area of the substrate (SiC in the present case); $Q$ is the vertical momentum transfer as in Figure \ref{fig:exp_md}a. Due to the loss of phase information inherent in the XR measurement, it is not possible to determine $\rho(z)$ by inverse FT. Instead, it is determined by optimizing a model wherein the $j^{th}$ atomic layer along the $z$-direction is represented by a Gaussian with fitting parameters of position, width, and coverage as described in the main text Eq. \eqref{eq:rho_e}. The parameter values and their uncertainties are determined via non-linear least squares fitting following the Levenberg-Marquardt algorithm until the level of agreement (Eq. \eqref{eq:chi2}) between the calculated model-dependent reflectivity, $R(Q)_{calc}$, and the experimental data, $R(Q)_{exp}$, converges.

\subsection{DIW/EG/SiC Interface Structure}
\begin{figure}[!hb]
    \includegraphics[width=8.5cm]{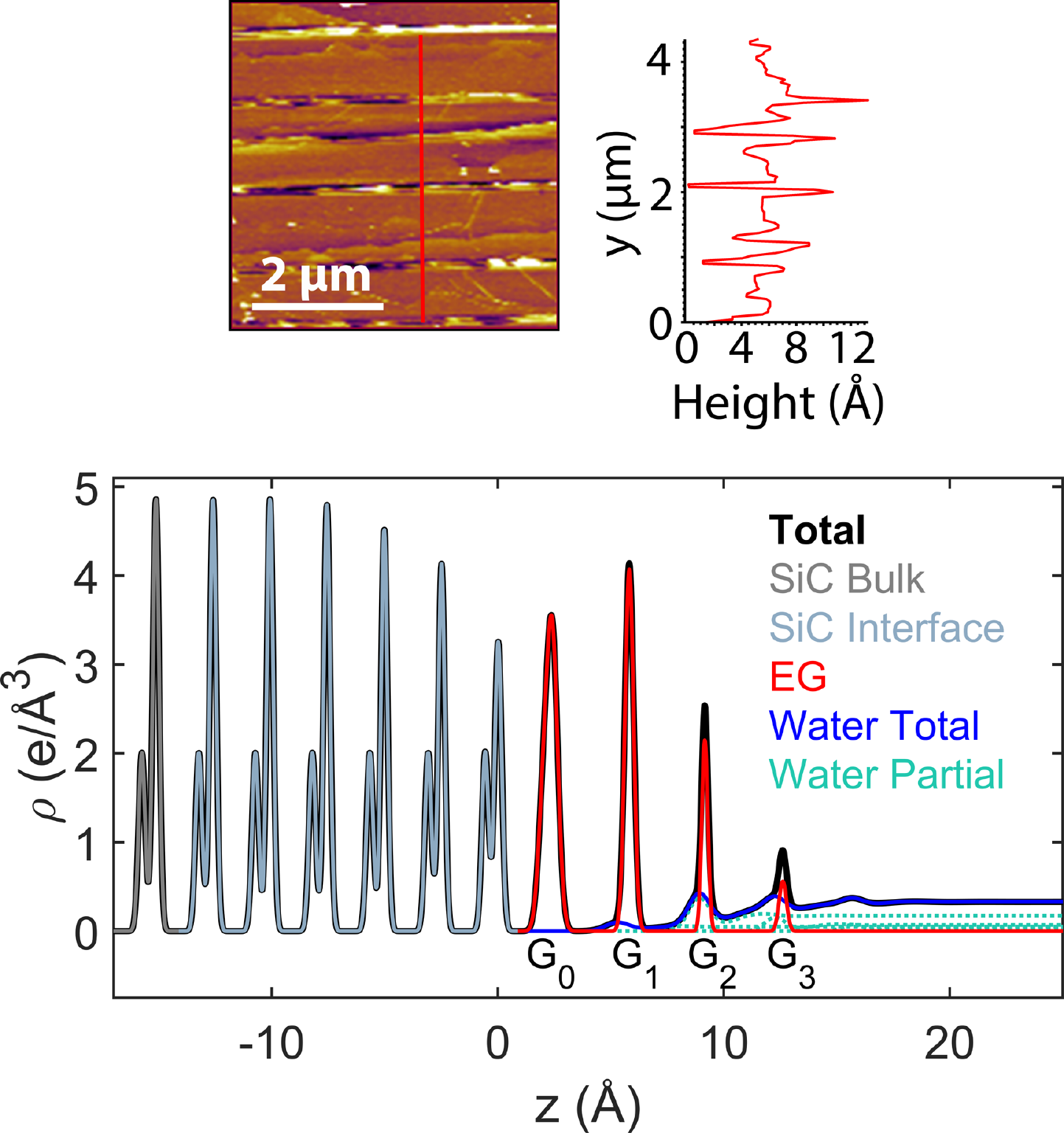}
    \caption{Top: AFM height image of EG/SiC sample with the line profile (red line) showing the step widths $y$ and heights; sharp spikes in the height profile indicate the step edges. Bottom: resolution-broadened best fit electron density profile from the CTR data with $\chi^2 = 1.6$ (Figure 2a in main text), including SiC substrate results and the partial water layers adsorbed on each exposed surface; $u_{broad} = (u_{i}^2+u_{res}^2)^{1/2}$ where $u_{res}=0.55/Q_{max}$.}
    \label{fig:SIexp}
\end{figure}

Figure \ref{fig:SIexp} shows an AFM height scan of the EG/SiC sample in air. It reveals a smooth sample surface (average surface roughness = 38 pm) with lateral terrace widths of $\sim1 \mu$m. We also see several layers of graphene on each terrace with a mean thickness of about 6 {\AA} and a maximum thickness of about 12 {\AA} (Figure \ref{fig:SIexp}, top). Due to the limited resolution of the AFM, it's chemical insensitivity, and the fact that it does not see the layers below the surface, we cannot determine the number of graphene layers precisely from the AFM. Nevertheless, this qualitative picture provides an additional reference point against which to evaluate the XR best fit structure (Figures \ref{fig2}a,b and \ref{fig:SIexp}, bottom). 

The optimized XR model parameters ($z$, $u$, and $\Theta$) for each atomic layer are summarized in Table \ref{Table:Fitparams}, and the complete best fit interface structure of de-ionized water (DIW) on EG/SiC is shown in Figure \ref{fig:SIexp}. We partially constrained the surface SiC and G$_0$ parameters in the XR analysis based on the work of Emery \textit{et al.}~\cite{Emery2013}. That study combined the chemical and structural sensitivity of X-ray photoelectron spectroscopy (XPS), X-ray standing waves (XSW), and XR to rigorously characterize the structure of the air/EG/SiC interface and address long-standing debates about the buffer layer, G$_0$, between SiC and epitaxial graphene. The XPS/XSW measurements were consistent with a carbon-rich buffer layer composed of two chemically-distinct and partially overlapping layers, S1 with sp$^2$ hybridization and S2 with sp$^3$ hybridization and bonded to Si atoms of the substrate~\cite{emtsev2007,Mattausch2007,Varchon2007,Varchon2008,Kim2008}. The results ruled out a proposed Si adatom model \cite{Rutter2007,Hass2008}. Although Emery \textit{et al.} found that the EG/SiC interface was largely identical across all samples studied (UHV-grown vs. Ar-grown with 1.3 or 1.7 ML graphene), we allowed for the structure parameters in this work to vary up to 10{\%} from their results. In general, our results are in agreement with those previously reported.

\subsubsection{G$_0$ Buffer Layer}
We identified a G$_0$ layer with a mean height of 2.31 $\pm$ 0.02 {\AA} above the SiC surface and FWHM of approximately 0.72 {\AA}. Similarly, Emery \textit{et al.} found a mean G$_0$ height of $\sim$2.3 {\AA} with a G$_0$ FWHM of $\sim$0.8-0.9 {\AA}. Others have reported G$_0$ heights of $\sim$ 2.5 {\AA} ~\cite{Mattausch2007,Rutter2007}. We note that although our XR measurement could not sufficiently resolve the two G$_0$ subpeaks S1 and S2, we were unable to obtain a good fit to the data using a single G$_0$ peak, likely due to the asymmetry of this layer. The resolution of the XR measurement is given by $r = \pi/Q_{max}$; $Q_{max}=4.911$ {\AA}$^{-1}$ resulting in $r\approx 0.64$ {\AA} for the current measurement, which is larger than the S1-to-S2 separation we identified of 0.42 $\pm$ 0.05 {\AA} but smaller than the FWHM of the combined G$_0$ layer. The S1-to-S2 peak separation is in agreement with the 0.35 {\AA} spacing reported by Emery \textit{et al.} We found that the spacing between S1 and S2 was conserved throughout the fitting iterations as both layers moved together with respect to the SiC surface, lending support for the shape of the buffer layer. Moreover, the distance from S2 to the topmost Si layer of the substrate was 1.97 $\pm$ 0.05 {\AA}, in agreement with earlier reports~\cite{Mattausch2007,Emery2013}. 
We find a G$_0$ coverage equivalent to 1.18 $\pm$ 0.03 ML, in contrast to previous results that identified a layer with essentially the density of graphene~\cite{Emery2013,Kim2008,Zhou2012}. Attempts were made to constrain the G$_0$ density to that of a single graphene layer, but such a density was always found to be inconsistent with the data. The excess carbon density we identified in the G$_0$ layer may account for a surface oxide species not included in our model~\cite{Emery2013,Bernhardt1999,HOSHINO2002234,Amy4342,Amy165323}. Emery \textit{et al.} identified via XPS the presence of SiO$_{\rm{x}}$ in several EG/SiC samples, which they were also unable to accurately model in the XR data analysis. They estimated the oxygen coverage to vary from 2 O/nm$^2$ in an Ar-grown sample to 6 O/nm$^2$ in a UHV-grown sample. The excess carbon density of the G$_0$ layer in our Ar-grown sample can equivalently be attributed to an oxygen content of $\sim$5 O/nm$^2$.

\subsubsection{SiC Surface}
We identify a partially depleted SiC surface, consistent with the thermal desorption of Si during graphene growth~\cite{VANBOMMEL1975463,Forbeaux1998,Huang2008}. We find that Si was depleted down to the fourth surface layer of the SiC while the C layers within SiC were not depleted. The topmost Si layer was displaced away from the bulk and toward the G$_0$ buffer layer. The coverage of the topmost Si layer is consistent with the coverage of the S2 layer of G$_0$ (though we report a large uncertainty on the S2 coverage, see Table \ref{Table:Fitparams}). We were unable to rigorously quantify the S2 layer coverage due to the large covariance of this parameter with the S1 coverage as a result of the limited resolution of the XR measurement discussed previously. Therefore, we fixed this parameter after several fitting iterations where it was converging to values consistent with the amount of surface Si depletion. That is, the coverages of the two layers indicate a one-to-one bonding between dangling Si atoms and sp$^3$-hybridized carbons, consistent with the proposed growth mechanisms and previous reports ~\cite{VANBOMMEL1975463,Forbeaux1998,Huang2008,Emery2013}. 

\subsubsection{Epitaxial Graphene}
We find a graphene film structure that is consistent with the AFM in Figure \ref{fig:SIexp}. Namely, we identify three 2D graphene layers, G$_1$ - G$_3$, above the G$_0$ buffer layer for a total of 1.25 $\pm$ 0.07 ML of epitaxial graphene. The mean graphene layer spacing is 3.40 $\pm$ 0.07 {\AA}, in agreement with the known value~\cite{Baskin1955}, and G$_1$ is located 3.5 $\pm$ 0.04 {\AA} above G$_0$, in agreement with previous reports~\cite{Emery2013,Kim2008,Varchon2008}. 

\subsubsection{Adsorbed Water}
The overall water structure results are discussed in the main text, and the best fit parameter values are given in Table \ref{Table:Fitparams}. Here, we focus on similarities and differences between our results and a those from a previous XR study of the adsorbed water structure on graphene \cite{Zhou2012} and discuss the implications for the intrinsic interactions of water with graphene. 

We find a water height above the free standing graphene layers (G$_1$-G$_3$) of $\approx3.1$ {\AA}, in agreement with the results of Zhou \textit{et al.} \cite{Zhou2012}. However, we identify significant differences in the water structure above the G$_0$ buffer layer. We find that water adsorbs closer to the buffer layer at $\sim$ 3 {\AA} above G$_0$, but the height uncertainty overlaps with the water height above free standing graphene and indicates that the buffer layer is also weakly hydrophobic. We note that although we assumed a single water model above all "graphene" surfaces in our analysis (including G$_0$), the different height for the buffer layer results from calculating the distance between water and the weighted average position of the S1 and S2 peaks, a condition that was not imposed during the XR data analysis. In contrast, Zhou \textit{et al.} found that water adsorbs at a height of $\approx$ 2.33 {\AA} above the buffer layer, indicating a hydrophilic character. They report water contact angle (WCA) measurements that support a more hydrophilic G$_0$ and which are equivalent to the WCA on bare SiC \cite{ShinDefect2010}. The WCA measurements showed a linearly increasing trend with graphene layer thickness, which may initially suggest that the buffer layer and subsequent graphene layers display wetting transparency on SiC. However, the extent of graphene's wetting transparency is still contested \cite{rafiee2012wetting,Raj2013,Shih2012,ShinDefect2010}, and it is unclear if the G$_0$ layer with its sp$^2$ and sp$^3$ character would exhibit similar wetting transparency properties. Instead, the evidence presented by Zhou \textit{et al.} suggest that the G$_0$ water adsorption is more in line with a high concentration of defects on their samples.

Zhou \textit{et al.} used UHV-grown EG/SiC samples, which are known to possess a greater amount of defects than graphene grown in a furnace in an Ar atmosphere (as was done for the sample studied in this work), as shown by their AFM images and other reports~\cite{EmtsevEGSiC}. Based on the results of Emery \textit{et al.} wherein UHV-grown and Ar-grown EG/SiC were found to have equivalent interface structures \cite{Emery2013}, which are consistent with our own EG/SiC interface structure, we would not expect the growth methodology to substantially contribute to the differences observed between our and Zhou's G$_0$-water distance. However, the quality of the sample depends on the vacuum level and any pre-treatments of the SiC to remove oxides \cite{Bernhardt1999,HOSHINO2002234,Amy165323,Seyller2006}. No pre-treatments were reported by Zhou \textit{et al.} They also reported Raman data with significant D and D+D' peaks, which result from edge and other defect states~\cite{Tuinstra,Ferrari2013,Ni2008,Casiraghi2009,ShinDefect2010}. In fact, it has been shown that the introduction of such defect peaks upon oxygen plasma etching of EG/SiC is associated with a decrease in WCA \cite{ShinDefect2010}. Finally, Zhou \textit{et al.} report MD and ab initio MD (AIMD) simulations of defect-free surfaces that predict water heights above G$_0$ consistent with that observed above free-standing graphene, and in agreement with our MD results. The AIMD simulations included effects of the SiC substrate and the corrugation of the buffer layer, but found only a $\sim$0.2 {\AA} decrease in the adsorbed water height above G$_0$ compared to free-standing graphene. Only upon inclusion of Si vacancies and -OH defects were they able to simulate a G$_0$-water height of 2.33 {\AA}. We conclude that while we can reasonably expect water to adsorb more closely to the buffer layer than to the subsequent graphene layers as a result of the SiC substrate and corrugated surface, we expect the effect to be minor in the absence of substantial defects. For a defect-free surface, the water structures above the buffer layer and subsequent graphene layers are very similar and can be described well by a single water model given certain resolution limits of the XR measurement.

\begin{table}
\caption{XR best fit results with uncertainties on the last significant figures in parentheses. Values without uncertainties were fixed during analysis.}
\begin{center}
\begin{tabular}{p{2cm} p{2cm} p{1.5cm} p{2cm}}
   \hline\hline
   \textbf{Layer} & \textbf{$z$ ({\AA})} & \textbf{\textit{u}({\AA})} & \bm{$\Theta (A_{\rm{UC}}^{-1})$}\\
   \hline
   \multicolumn{4}{c}{SiC}\\
   \hline
   C & -13.226(5) & 0.0922 & 1 \\
   Si & -12.598(2) & 0.0837 & 1\\
   C & -10.723(7) & 0.0922 & 1 \\
   Si & -10.076(3) & 0.0837 & 1\\
   C & -8.204(15) & 0.0922 & 1 \\
   Si & -7.557(5) & 0.0837 & 0.988(8)\\
   C & -5.662(17) & 0.0922 & 1 \\
   Si & -5.026(5) & 0.0837 & 0.930(21)\\
   C & -3.080(22) & 0.0922 & 1 \\
   Si & -2.488(10) & 0.0837 & 0.850(34)\\
   C & -0.555(73) & 0.0922 & 1(1)\footnotemark[1]\\
   Si & 0.019(15) & 0.11(13)\footnotemark[1] & 0.748(43)\\
   \hline
   \multicolumn{4}{c}{Graphene}\\
   \hline
   S1 (G$_0$) & 2.41(3) & 0.224(25) & 2.86(11)\\
   S2 (G$_0$) & 1.99(5) & 0.19(3) & 0.86(44)\footnotemark[1]\\
   G$_1$ & 5.82(4) & 0.154(22) & 2.66(12)\\
   G$_2$ & 9.16(7) & 0.07(27)\footnotemark[1] & 0.98(18)\\
   G$_3$ & 12.6(1) & 0.12(71)\footnotemark[1] & 0.32(3)\\
   \hline
   \multicolumn{4}{c}{Water}\\
   \hline
   H$_2$O & 5.29(34) & 0.43(37) & 0.66\footnotemark[2]\\
   & $d_w$ = 2.42(1.46)\footnotemark[1] & $\bar{u}=$ 1(1)\footnotemark[1]&\\
   \hline\hline
\end{tabular}
\footnotetext[1]{Large uncertainties indicate a general insensitivity to these structural features and magnify uncertainties for other parameters in cases where they covary. As such, these parameters were fixed in the final iterations of the least-squares optimization.} \footnotetext[2]{Calculated from $d_w$.}
\label{Table:Fitparams}
\end{center}
\end{table}

\subsection{Agreement and Differences with Simulation}
\begin{figure}[ht]
\includegraphics[width=7.5cm, trim={0 0.35 0 0.5cm},clip]{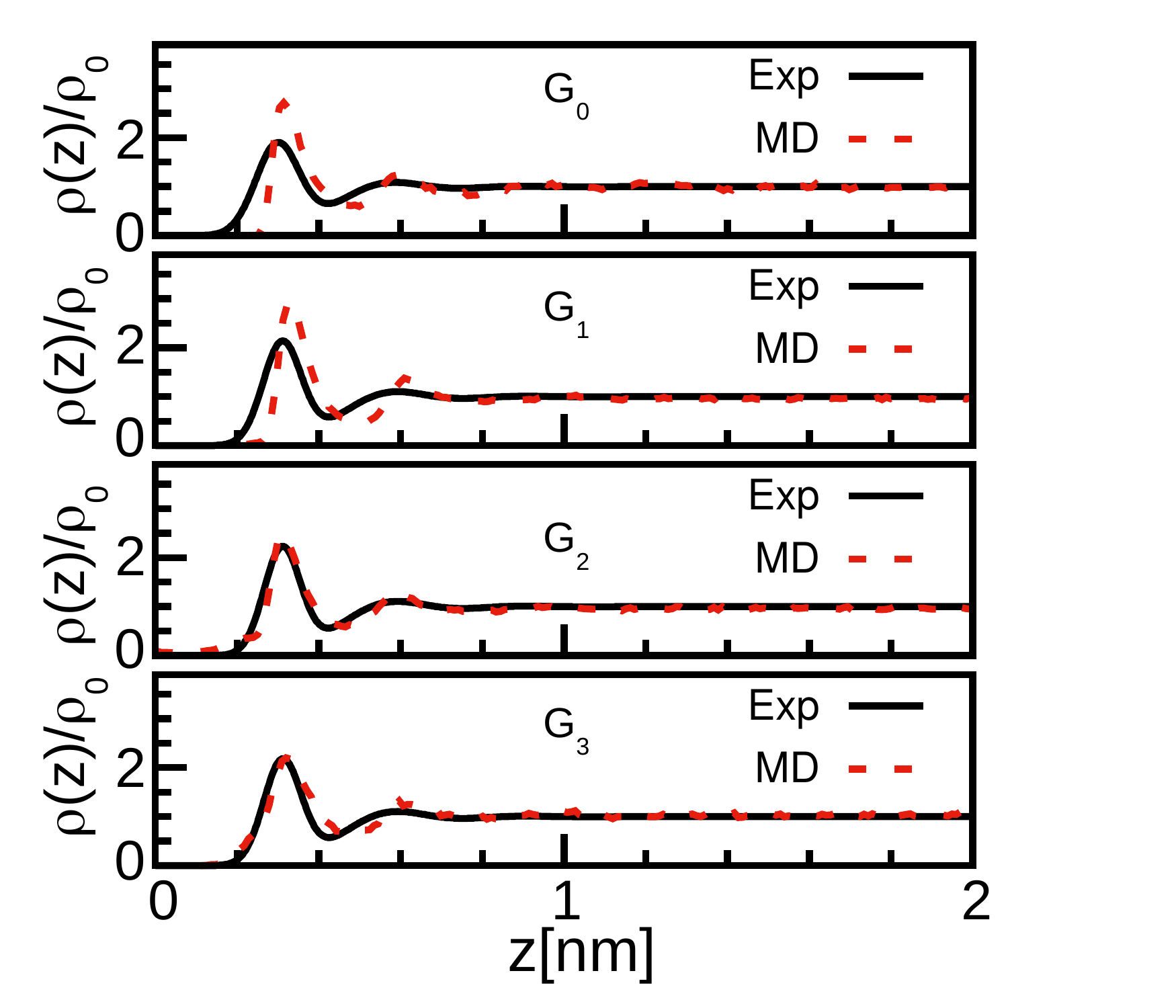}
\caption{Density profiles normalized with respect to the bulk water density on top of the layer G$_i$ ($i=0,\dots, 3$) from experiments (solid black line) and from MD simulations (dashed red line). The widths of the hydration layers in the experiment are broadened according to Eq.\eqref{eqn8}.}
\label{fig:MDXRwater}
\end{figure} 
Figure \ref{fig:MDXRwater} shows adsorbed water profiles isolated from each exposed graphene layer (see Figure 2d in the main text for the \textit{total} density profile). Although the XR and MD results generally agree, a noticeable broadening of the first hydration layer is predicted by MD above each partial graphene surface. The width is consistent with the XR result by the third layer (G$_2$). A graphene slab of four uniform and complete layers produces the same water structure as is shown in Figure \ref{fig:MDXRwater} for the layer G$_0$. The origins of these discrepancies should be discussed in the context of the limitations of both the MD and XR approaches:\newline

(1) The G$_0$ layer of the MD simulation is modeled as a 2D graphene layer due to the absence of a SiC substrate. However, the partial sp$^2$ and sp$^3$ character of G$_0$ on SiC alters the band structure~\cite{Ohta_ARPES} and may affect the buffer layer's interaction with water. Indeed, this is consistent with our XR results and previous AIMD predictions \cite{Zhou2012}, though the effect is weak. Previous MD simulations of unsupported graphene sheets suggest via water contact angle calculations that the graphene hydrophobicity decreases with increasing layer thickness  \cite{Shih2012} and becomes graphite-like by the third layer \cite{Taherian}. This agrees with our MD simulations wherein the adsorbed water peak height is conserved, but the broadening reveals a small increase in the number of water molecules that are able to adsorb closer to the graphene sheet. \newline\newline
(2)The XR-derived result is limited by the complexity of the XR analysis model, as described in detail above. We used a single \textit{intrinsic} water model for each graphene surface despite evidence that on SiC thicker graphene regions are more hydrophobic than thinner regions \cite{Kazakova2015}. We note that this is the opposite of MD predictions, suggesting that while the effect of the SiC substrate and corrugated buffer layer are relatively weak in magnitude, they may significantly affect the chemistry of the surface. Accounting for different hydrophobicity would introduce additional parameters to an already complex model and is unnecessary given that the changes are small. Using a single water model essentially gives an optimized structure that captures the average behavior of water adsorption on an imperfect graphene surface. In addition, the non-linear least-squares fitting of the data can produce multiple equally viable structures (i.e., with equivalently good $\chi^2$). This point emphasizes the importance of including evidence from other experimental methodologies such as XPS/XSW \cite{Emery2013} to refine and constrain models. 
Even with these caveats, the qualitative agreement between the XR and MD structures, paired with the high-confidence in the XR result (given a $\chi^2$ of 1.6 where a perfect fit would have a $\chi^2$ of 1), provides a consistent picture of water adsorption on graphene.


%

\end{document}